\documentclass[11pt]{article}

\usepackage[english]{babel}

\usepackage[letterpaper,top=1.8cm,bottom=1.8cm,left=3cm,right=3cm,marginparwidth=1.75cm]{geometry}

\usepackage{amsmath}
\usepackage{booktabs}    
\usepackage{graphicx}
\usepackage{tabularx}
\usepackage[colorlinks=true, allcolors=blue]{hyperref}
\usepackage{setspace}
\usepackage{natbib}
\usepackage{amsthm}
\usepackage{booktabs}
\usepackage{multirow}
\usepackage{array}
\usepackage{amsmath}
\usepackage{amssymb}
\usepackage[title]{appendix}
\usepackage{caption}
\usepackage[perpage]{footmisc}
\usepackage{hyperref}
\setcitestyle{authoryear,round}
\newcolumntype{L}[1]{>{\raggedright\arraybackslash}p{#1}}
\newcolumntype{C}[1]{>{\centering\arraybackslash}p{#1}}

\newtheorem{Hypothesis}{Hypothesis}

\title{\textbf{The Nexus between Dataization and Technological Progress in General Equilibrium of Macroeconomics}}
\author{Yongheng Hu\thanks{School of International Business, Zhejiang International Studies University, Liuhe Road, Hangzhou 310023, China. Correspondence to: Yongheng Hu (22030101043@st.zisu.edu.cn). Previous versions of this article have been accepted and publicly reported in several conferences: \textit{Big Data Industry-Academia-Research Summit Forum in 2024}. \textit{Undergraduate Forum on New Structural Economics in 2025}. \textit{International Congress of Basic Sciences (ICBS) in 2025}. \textit{Young Economists Forum of China in 2025}. I am very grateful to Dr.Mingyi Yang of Washington State University for his guidance and many helpful discussions. This working paper is incomplete and still being improved, all comments and opinions about the article are welcome. Of course all remaining omissions and errors are mine.}}

\begin{document}
\maketitle

\begin{abstract}
In this paper, we construct an analytical model of the data economy with empirical evidence to explain the nexus between dataization and technological progress in general equilibrium. Data originates from the dataization of firms’ total output and contributes to the formation and enhancement of technology. Firms use the production function with data to solve the optimal investment, while households use the endogenous interest rate from the firms’ problem to solve the optimal consumption. We find that dataization has a negative moderating effect on the transition of general equilibrium affected by technological progress. Policy can only facilitate a positive transition in general equilibrium by simultaneously encouraging dataization and technological progress. Furthermore, when equilibrium capital stock is in a stationary state, dataization enhances technological progress at high levels. However, when equilibrium consumption is in a stationary state, dataization enhances technological progress at low levels while weakening it at high levels. Our empirical analysis uses macroeconomic data and policy from Chinese cities during 2000–2021 to verify the theories proposed in this paper. We further apply the Mean Field Games in a continuous-time framework to provide an extended explanation for the nexus between dataization and technological progress in partial equilibrium.


\end{abstract}

\newpage

\section{Introduction}

Data essentially shifts the material basis of production from traditional discourse practices to quantifiable and tradable data resources, completing the dataization of knowledge, information and traditional production factors. The special economic attributes of data as a kind of new production factor, such as non-rivalry, replicability, virtuality, open source, and non-depletion at the physical level (\citealp{jones2020nonrivalry}; \citealp{goldfarb2019digital}; \citealp{lynch2008how}), enable it to be integrated into the economic cycle of production by decreasing marginal costs and increasing returns to scale.

Many current researches on the impact of data on economic systems focus on how data factors improve productivity and promote economic growth, exploring the role of data that empower technological progress in terms of productivity enhancement, especially the total factor productivity and the reduction effect on various production costs (\citealp{nose2023inclusive}; \citealp{acemoglu2018race}; \citealp{graetz2018robots}), as well as treating data as intangible capital and constructing economic models with data capital to study the impact of big data on productivity \citep{corrado2024data}. Moreover, data factors can effectively drive the innovation of digital products \citep{athey2025presidential}, since researching more novel problems can contribute to the evolution of knowledge, thus triggering the research cycle \citep{carnehl2025quest}, AI technologies empowered by data factors can significantly increase the amount and accuracy of scientific knowledge acquisition in a specific domain,  as a result, the desire of researchers to innovate and the efficiency of technological innovation can be effectively improved, then, researches on more novel problems will be facilitated (\citealp{gans2025quest}; \citealp{costa2025exploring}). Meanwhile, the application of data factors can also play a role in driving the transformation of high-quality industries and optimizing the allocation of production resources (\citealp{acemoglu2025simple}; \citealp{glaeser2018big}; \citealp{zhang2018survey}; \citealp{schaefer2014long}; \citealp{einav2014economics}). 

There are also studies of data-driven productivity based on the microeconomic attributes of data factors. As a result of the revolution in production by digital technology, data factors are transformed more into the knowledge for production process, which improves the production chain, regulates the rationing of production factors, drives technological innovations and raises the marginal productivity of the traditional factors \citep{abis2024changing}. Through its own “multiplier effect”, data can empower traditional production factors and contribute to capital-expanding and labor-expanding technological advances individually or simultaneously, thus enhancing the marginal productivity of capital and labor factors (\citealp{hemous2022rise}; \citealp{bessen2019automation}; \citealp{acemoglu2018race}).

Meanwhile, viewing data as a special kind of labor can help alleviate problems in terms of unequal returns to the data economy and restore a normal market for household contributions \citep{arrieta2018should}. However, the use of a “data” workforce is not always beneficial, and when firms overuse data-driven automation to eliminate traditional human capital, this leads to borrowing and redistributive frictions, which affects the efficiency of economic operations \citep{beraja2025inefficient}. Moreover, \citet{brynjolfsson2025generative}, the perspective of heterogeneous workers finds that the AI technology driven by big data can improve both productivity and quality of work for low-qualified workers, but for high-qualified workers, it only slightly improves the productivity of their work, while the quality of their work decreases. \citet{aum2025labor} show that digital technology enabled by big data has a significant negative impact on high-skilled workers, and while this can displace part of labor, it can also drive labor demand towards newer skill sets, thus boosting the total labor demand of economics \citep{hampole2025artificial}. However, \citet{jiang2025ai} suggest that the relationship between data-driven AI technologies and the human workforce is complementary rather than substitutive, and that will lead to a significant improvement in the marginal productivity of economics, and will serve as an effective proxy for leadership, decision-making and teamwork \citep{weidmann2025measuring}, but also disrupts the work-life balance of individuals.

In addition, \citet{farboodi2019big} and \citet{farboodi2021model} regard data as an information resource that can reduce production uncertainty, and firms use data elements to acquire forward-looking knowledge, thus improving the accuracy of technological prediction, which further improves the level of productivity. The findings of \citet{comin2018technology} and \citet{varian2018artificial} indicate that the diffusion of data factors in economics is similar to a logistic function, the location of diffusion inflection points and the upper limit of diffusion are determined by the diffusion intensity of data factors. Then, a number of studies on how to balance the dimensions of “efficient output” and “social equality or security” have pointed out that the depth and breadth of data applications have continued to increase with the rapid development of technology, which completely changes the economy, business, public administration, national security, scientific research, consumer rights and many other areas \citep{lazer2014parable}. Specifically, as the degree of big data sharing increases, it improves consumer surplus and consumer utility. However, at the same time, users buy the services from firms and contribute data, then firms invest in data infrastructure and collection. Hence, the data generated by users will be used for the competitions of firms through the data-driven model or algorithm. That is, data collection improves services and benefits users, but users also face security issues such as privacy breaches \citep{cong2023data}, which will incur privacy costs derived from data oversharing (\citealp{cong2021}; \citealp{jones2020nonrivalry}).

Similarly, online platforms can use consumer information materialized by data factors to alter the “gloss” of product quality, causing consumers to misjudge the true quality of the product and inducing irrational consumption behaviors \citep{acemoglu2025when}. Moreover, platforms will not only use the pricing mechanism called “kill maturity from big data” to make profits, but will also redistribute information in the form of “filter bubbles” from heterogeneous personal data to maximize platform engagement \citep{acemoglu2024model}. A study led by \citet{ding2024consumer} argues that the positive effect on consumption growth transformed from data factors, such as the issuance of digital consumption vouchers or shopping subsidies, comes from the increase of consumer’s spending in the targeted category, and is not due to the emergence of other universal categories of consumption expenditure. 

\citet{madsen2025insider} point out that a complete ban on data use may foster or inhibit innovation. Hence, more flexible policies that control what kind of data should be generated for dataization are needed to increase the effectiveness of data regulation because of the demand for new products. Some studies have incorporated information uncertainty into the “efficiency-equity” research framework, pointing out that information frictions connect data generation and economic activity by the following ways: Microeconomic uncertainty creates resource mismatches in the macroeconomics, making macro-level total factor productivity endogenous to micro-level agents’ data collection behavior (\citealp{farboodi2021model}; \citealp{david2016information}; \citealp{benhabib2016endogenous}). This may be due to the widespread application of facilitated mobile communication tools and the availability of data information that allows for more random and rapid changes in individual behavior \citep{fabregas2025digital}, which leads to the inability of traditional data selection mechanisms to accurately identify imperfect information and decision-making errors. As \citet{gans2025strategist} points out that in the domains with rich varieties of dataization, AI’s analytical and decision-making capabilities perform well, while it is less credible in the dataization with intensive judgment environments. Therefore, \citet{caplin2025data} introduces new forms of data to identify the preferences and beliefs of agents by constructing data engineering models.

Notably, \citet{jones2025how} and \citet{jones2024ai} explore the potential of AI technology enabled by big data to imply a threat to human existence while promoting economic growth by data factors according to the utility acquisition of heterogeneous agents. 

To summarize, there are two main kinds of studies in current economic research: The first perspective views the data as a form of medium for information dissemination, which determines the source and value of information \citep{farboodi2021model}. And the second perspective views data as an input factor for production, which determines technological progress and aggregate output \citep{jones2020nonrivalry}. The analysis of data economy in this paper refers to \citet{jones2020nonrivalry}. We regard data as a new kind of production factor to study its impact on macroeconomics\footnote{The framework of our paper is based on the perspective of “data economy”, which is different from the “digital economy”. Data economy regards data as a factor of production, which centers on the value creation through production, distribution, exchange and consumption of data resources. However, digital economy covers the two dimensions of digital industrialization and industrial digitization. It could be manifested in the expansion of production boundary or the innovation of supply chain and industrial organization, etc.}.

The motivation for this paper comes from our curiosity about how dataization and technological progress affect the capital stock and consumption in society. Hence, we aim to analyze how dataization and technological progress will influence the transitions in general equilibrium of macroeconomics. Meanwhile, we have already established that dataization and technological progress can jointly constitute part of the aggregate production function in the form of exponential functions, that is, we have clearly identified the connection between dataization, technological progress and production. However, we still have some uncertainties regarding the nexus between dataization and technological progress. We attempt to construct an analytical model to explore how this inherent relationship will affect the transitions in general equilibrium and identify the relevant economic reasons that sustain this relationship.

We apply macroeconomic data from China during 2000–2021 and design econometric research to explore how dataization policy and technological progress affect general equilibrium, i.e., social capital stock and social consumption. We also examine whether the nexus between dataization and technological progress aligns with the conclusions derived from our theoretical model. We then analyze the heterogeneity in the effects of dataization policy based on regional income and consumption inequality, providing potential insights and references for the formulation or refinement of future policies related to the data economy.

\section{Prerequisite Model}

\subsection{The Production Function with Data}
We assume that all consumers and firms in the macroeconomics of this paper are homogeneous, i.e., representative consumers and representative firms. In this macroeconomics, the consumption of representative consumers and the investment of representative firms will generate data, and these data factors can be allocated to the formation of algorithms or the advance of technology to improve production. 

Here, we assume that data is derived from the dataization of the total output in economy and applied to enhance technology. At the same time, the dataization proportion of total output is a policy parameter determined by the government. That is, policies do not allow some of the data to be applied to algorithms for studying consumer and business behavior due to the dual factors of enhancing consumer welfare and protecting individual privacy, or policies stipulate a certain percentage of the total output can be applied to the development of technology or innovation after dataization.

Hence, for the \textit{Cobb-Douglas} production function without data:
\begin{align}
y_0(t)=(z(t)k(t))^\alpha l(t)^\beta \label{(1)}
\end{align}

Where $y(t)$ is output, $z(t)$ is the technological attachment of capital, $k(t)$ is capital, and $l(t)$ is labor. $\alpha$ and $\beta$ are the output elasticities of capital and labor, respectively, with $\alpha \in (0,1)$ and $\beta \in (0,1)$.

Now, taking data as a factor of the production function, based on \citet{jones2020nonrivalry}, we assume an exponential relationship $\eta$ between data $d(t)$ and technology $z(t)$. Introducing the exogenous variable $\theta$, we set that the policy stipulates only $\theta$ proportions of total output could be applied and put into production as data factors after dataization, i.e.:
\[\begin{gathered}z(t)=d(t)^\eta, \eta \in (0,1)\\d(t)=\theta y(t), \theta \in (0,1)\end{gathered}\]

Solving by associating $y(t)$,$z(t)$,$d(t)$, we get:
\begin{align}
y(t)=(\theta^\eta k(t))^\frac{\alpha}{1-\alpha\eta}l(t)^\frac{\beta}{1-\alpha\eta}=\theta^\frac{\alpha\eta}{1-\alpha\eta}k(t)^\frac{\alpha}{1-\alpha\eta}l(t)^\frac{\beta}{1-\alpha\eta}\label{(2)}
\end{align}

In equation (2), $\alpha\in(0,1)$ and $\eta\in(0,1)$, thus we obtain:
\[\frac{\alpha}{1-\alpha\eta}+\frac{\beta}{1-\alpha\eta}=\frac{\alpha+\beta}{1-\alpha\eta}>\alpha+\beta\]

This clearly demonstrates that the returns to scale of an economy represented by a production function with data participation are always greater than those of a production function without data participation. That is, compared to production function (1), which does not incorporate data, the returns to scale of production function (2) in an economy with data participation will increase, and the degree of this increase depends on the multiplier $\alpha\eta$.

Furthermore, we find that for the production function (1), which does not incorporate data, it inevitably exhibits decreasing marginal productivity for both labor and capital, which is:
\[\begin{gathered}
\frac{\partial^2y_0(t)}{\partial l(t)^2}=\beta(\beta-1)(z(t)k(t))^\alpha l(t)^{\beta-2}<0,\beta\in(0,1)\\\frac{\partial^2y_0(t)}{\partial k(t)^2}=\alpha(\alpha-1)z(t)^\alpha k(t)^{\alpha-2}l(t)^\beta<0,\alpha\in(0,1)
\end{gathered}\]

However, for the production function (2) with data participation, it may exhibit increasing marginal productivity for both labor and capital, that is:
\[\begin{gathered}
\frac{\partial^2y(t)}{\partial l(t)^2}=\frac{\beta}{1-\alpha\eta}(\frac{\beta}{1-\alpha\eta}-1)(\theta^\eta)^{\frac{\alpha}{1-\alpha\eta}}k(t)^{\frac{\alpha}{1-\alpha\eta}}l(t)^{\frac{\beta}{1-\alpha\eta}-2}>0,\eta\in(\frac{1-\beta}{\alpha},1)\\\frac{\partial^2y(t)}{\partial k(t)^2}=\frac{\alpha}{1-\alpha\eta}(\frac{\alpha}{1-\alpha\eta}-1)(\theta^\eta)^{\frac{\alpha}{1-\alpha\eta}}k(t)^{\frac{\alpha}{1-\alpha\eta}-2}l(t)^{\frac{\beta}{1-\alpha\eta}}>0,\eta\in(\frac{1-\alpha}{\alpha},1)
\end{gathered}\]

If and only if:
\[\begin{gathered}\alpha\in\left(\frac{1}{2},1\right),\beta\in\left(\frac{1}{2},1\right)\\\eta\in(\frac{1-\beta}{\alpha},1)\cap(\frac{1-\alpha}{\alpha},1)\end{gathered}\]

At this time, we can simultaneously achieve increasing marginal productivity for both labor and capital according to equation (2).

\subsection{The Nexus between Dataization and Technological Progress}
It is important to note that the exponential transformation rate between data and technology stems from the nonrivalry and cumulative effects of data. As a factor of production, data can be shared indefinitely without loss, and its value grows exponentially with the scale of its application. In the production function with technology, the amount of data directly affects the production efficiency of technology through the exponential form: 
\[z(t)=d(t)^\eta\]
\[\eta=\left(\frac{\ln z(t)}{\ln d(t)}\right)\]

Hence, $\eta$ reflects the importance of data to technology. As the amount of data increases, the output efficiency of technology increases at an exponential rate, forming a positive cycle: more data → higher technological efficiency → more output → generate more data. 

For example, the training of autonomous driving techniques requires a large amount of driving data. Assuming that an algorithm uses 1 million kilometers of driving data with $80\%$ accuracy, when the amount of data is increased to 10 million kilometers (10 times), the accuracy may be increased to $95\%$ due to the cumulative effect of nonrival data (nonlinear growth). This improvement is not linearly cumulative, but rather the data is trained by the algorithm to form an exponential optimization, reflecting the exponential transformation relationship between data and technology.

Now, we analyze the nexus between dataization and technological progress. We define aggregate output $Y$ in the macroeconomics of this paper as the combination of partial outputs $y(u)$ from each distinct task $u\in(0, 1)$. That is, all tasks based on $u$ are combined via a \textit{Cobb-Douglas} aggregator:
\[\ln Y=\int_0^1\ln y(u)du\]

We assume that the production of task $u$ requires the allocation of effective labor and effective capital in the \textit{Leontief} form. That is, in completing each specific task $u$, we only use capital or labor to finish that task according to the principle of optimal resource allocation:
\[y(u)=\min\{L(u),K(u)\}\]

Then, under optimal conditions:
\[L(u)=K(u)=y(u)\]

In this section, we will focus solely on the economic benefits arising from the incorporation of data in the labor factor. On the one hand, since the mathematical status of labor and capital factors is isomorphic in the \textit{Cobb-Douglas} production function, we can use the same methods and procedures to analyze the impact from the incorporation of data in the capital factor on total factor productivity. On the other hand, compared to the relationship between the capital after dataization and traditional capital, the labor after dataization is more likely to form a perfect and complete substitution relationship with traditional labor. For example, the widespread adoption of industrial robots is essentially a combination of production data and traditional capital, which has led to job displacement among workers.

Referencing the framework of \cite{moll2022uneven}, we define the effective labor factor $L(u)$ as consisting entirely of the traditional labor factor $l(u)$ and the labor factor after dataization $d_{L}(u)$. That is:
\[L(u)=l(u)+\psi_L(u)d_L(u)\]

Here, $l(u)>0$ and $d_{L}(u)>0$. $\psi_{L}(u)$ represents the efficiency of data substituting for labor in task $u$. We assume that continuous tasks $u\in(0,1)$ are ordered by the difficulty of substitution. Following the principle of maximizing production efficiency, we will first prioritize dataization for tasks where $u\rightarrow0$. This implies that, for task $u$, the earlier dataization is completed, the stronger the necessity for the data production factor to replace traditional production factors, and the higher the efficiency of the data production factor in substituting traditional production factors:
\[\psi_L(u)=\bar{\psi}_L\phi(u)\]

Where $\bar{\psi}_{L}>0$ and $\phi^{'}(u)<0$. Then, we simply define $\phi(u)$ as $\phi(u)=1-u$.

Furthermore, we assume that the maximum proportion of labor that can be replaced by data is $\hat{a}\in(0,1)$. That is, for the output $y(u)$ of a continuous task $u\in(0,1)$, at most $\hat{a}\in(0,1)$ of the output is produced by the labor factor after dataization $d_{L}(u)$, while the remaining $(1-\hat{a})\in(0,1)$ is produced by the traditional labor factor $l(u)$.

Hence:
\[\frac{\psi_L(u)d_L(u)}{l(u)}\leqslant\frac{\hat{a}}{1-\hat{a}}\]

That is to say, when the substitution proportion reaches the maximum:
\[\begin{gathered}
\psi_L(u)d_L(u)=\hat{a}y(u)\\l(u)=(1-\hat{a})y(u)
\end{gathered}\]

For the government, as described in Section 2.1, we assume that firms obtain data from the dataization proportion $\theta\in(0,1)$ of their total output, and $\theta$ is subject to government regulation. Therefore, we assume that the government permits the total amount of data used in production not to exceed a proportion $\theta\in(0,1)$ of firms’ total output, which is:
\[\int_0^1d_L(u)du\leqslant\theta Y\]

Next, the firm determines the amount of data to allocate to each task, aiming to maximize aggregate output $Y$ subject to the constraint of the total data supply. Since $\psi_{L}(u)$ decreases as $u$ increases within $(0,1)$, data should be prioritized for tasks with smaller $u$, ensuring these tasks reach the substitution threshold until data is exhausted. We set a threshold $u_{L}\in[0,1]$. All tasks with $u\in(0,u_{L})$ can reach the data substitution threshold, and are completed jointly by data and labor. Meanwhile, all tasks with $u\in(u_{L},1)$ do not use data and must be completed entirely by labor alone.

Therefore, for $u\in(0,u_{L})$:
\[\begin{gathered}
l(u)=(1-\hat{a})y(u)\\d_L(u)=\frac{\hat{a}}{\psi_L(u)}y(u)
\end{gathered}\]

And for $u\in(u_{L},1)$:
\[\begin{gathered}
l(u)=y(u)\\d_L(u)=0
\end{gathered}\]

Since each task for every $u\in(0,1)$ is symmetric, and aggregate output (final good production) follows a \textit{Cobb-Douglas} aggregator, according to \textit{Jensen's} inequality:
\[\ln Y=\int_0^1\ln y(u)du\leqslant\ln\int_0^1y(u)du\]

The equality holds if and only if $y(u)$ is a constant function of $u$, which also implies that the outputs of all tasks are exactly equal. Therefore, in the optimal scenario, we get $y(u)=Y$. And the aggregate production from the traditional labor factor $\mathcal{L}_0$ is:
\[\mathcal{L}_0=\int_0^{u_L}(1-\hat{a})Ydu+\int_{u_L}^1Ydu=Y(1-\hat{a}u_L)\]

Meanwhile, we can also obtain the aggregate production from the labor factor after dataization $\mathcal{L}(u)$:

\[\mathcal{L}(u)=y(u)=Y=\frac{\mathcal{L}_0}{1-\hat{a}u_L}\]

And the aggregate production from the data factor $\mathcal{D}$ is:

\[\mathcal{D}=\int_0^{u_L}\frac{\hat{a}}{\psi_L(u)}Ydu=\frac{\hat{a}Y}{\bar{\psi}_L}\int_0^{u_L}\frac{1}{\phi(u)}du=\frac{\hat{a}Y}{\bar{\psi}_L}\int_0^{u_L}\frac{1}{1-u}du=\frac{\hat{a}Y}{\bar{\psi}_L}\left[-\ln(1-u_L)\right]\]

According to $\int_{0}^{1}d_{L}(u)du\leqslant\theta Y$, we can solve and get:
\[\begin{gathered}
\frac{\hat{a}}{\bar{\psi}_L}[-\ln(1-u_L)]=\theta\\u_L=1-e^{\left(\frac{-\overline{\psi}_L}{\hat{a}}\right)\theta}
\end{gathered}\]

Then, based on equation (1), before introducing data, we treat the macroeconomics in this paper as a standard \textit{Cobb–Douglas} production function:
\[\mathcal{Y}=\mathcal{\bar{A}}\mathcal{L}^{\hat{\beta}}_0\mathcal{K}^{\hat{\alpha}}_0\]

Where $\mathcal{\bar{A}}=1$ represents traditional total factor productivity, $\mathcal{L}_0$ represents the traditional labor factor, and $\mathcal{K}_0$ represents the traditional capital factor.

We consider both traditional factors and total factor productivity as constants. After introducing data, the traditional capital factor remains unchanged, while data is incorporated into the traditional labor factor. At this scenario, total factor productivity $\mathcal{A}(\theta)$ will increase in production function $\mathcal{Y}(u)$:
\[\begin{gathered}
\mathcal{Y}(u)=\mathcal{\bar{A}}\mathcal{L}(u)^{\hat{\beta}}\mathcal{K}^{\hat{\alpha}}_0\\\mathcal{A}(\theta)=\mathcal{\bar{A}}(1-\hat{a}u_L)^{-\hat{\beta}}\
\end{gathered}\]

Substituting $u_L$:
\[\mathcal{A}(\theta)=\bar{\mathcal{A}}\left[(1-\hat{a})+\hat{a}e^{\left(\frac{-\bar{\psi}_L}{\hat{a}}\right)\theta}\right]^{-\hat{\beta}}\]

Now, we begin to analyze the relationship between data and technological progress. In Section 2.1, $\eta\in(0,1)$ represents the data-technology transformation rate, reflecting the exponential boost in technological levels resulting from the accumulation of data. However, in this section, data exhibits varying substitution efficiency across different tasks, which means that the data-technology transformation rate is no longer constant in aggregate production. Therefore, in a heterogeneous task environment, $\eta$ shifts from a fixed parameter to a variable that depends on the degree of dataization $\theta$, and the shape of the $\eta(\theta)$ curve reflects the nonlinear boundaries of technological progress driven by dataization policies. We define technological progress $\eta(\theta)$ as the elasticity of total factor productivity (technology) $\mathcal{A}(\theta)$ with respect to the degree of dataization $\theta$, i.e., the percentage change in total factor productivity for every $1\%$ increase in the degree of dataization:

\[\eta(\theta)=\frac{d\mathrm{ln}\mathcal{A}(\theta)}{d\mathrm{ln}\theta}=\theta\frac{d\mathrm{ln}\mathcal{A}(\theta)}{d\theta}\]

Finally, we get:
\[\eta(\theta)=\frac{\theta\hat{\beta}\hat{a}\left(\frac{\bar{\psi}_L}{\hat{a}}\right)e^{\left(\frac{-\bar{\psi}_L}{\hat{a}}\right)\theta}}{(1-\hat{a})+\hat{a}e^{\left(\frac{-\bar{\psi}_L}{\hat{a}}\right)\theta}}\]

We then perform a numerical simulation of $\eta(\theta)$. We define $\theta\in(0,1)$, and for the function $\eta(\theta)$, there exists a threshold $\theta^{*}$ which satisfies the following conditions:
\[\frac{d\eta(\theta<\theta^*)}{d\theta}>0,\frac{d\eta(\theta>\theta^*)}{d\theta}<0\]

We define three types of labor with maximum data substitution proportions $\hat{a}\in(0,1)$ are denoted as $(\hat{a}^{-},\hat{a}^{0},\hat{a}^{+})=(0.2,0.4,0.6)$. The output elasticity of labor $\hat{\beta}=0.2$. The baseline data substitution efficiency $\bar{\psi}_{L}=1$. The simulation results are shown in Figure 1:

\begin{figure}[htbp]
\centering
\includegraphics[width=15.5cm]{B1.jpg}
\caption{\label{fig:B1}The Nexus between $\theta$ and $\eta$ with Different $\hat{a}$}
\end{figure}

As shown in Figure 1, we observe that regardless of the value of the maximum proportion of labor that can be replaced by data, $\eta(\theta)$ exhibits an inverted U-shaped trend overall. Then, as $\hat{a}$ increases, the maximum value of $\eta(\theta)$ increases, and the threshold for the inverted U-shaped trend $\theta^{*}$ also increases.

\section{General Equilibrium}

\subsection{The Firm Problem}
\textbf{\textit{Endogenous Interest Rate}}. First, we consider the accounting profit of the firm and then we consider the investment based on the accounting profit of the firm. Accounting profit is expressed as:
\begin{align}
\pi(k,w)=\max_l\left\{\theta^\frac{\alpha\eta}{1-\alpha\eta}k^\frac{\alpha}{1-\alpha\eta}l^\frac{\beta}{1-\alpha\eta}-wl\right\}\label{(3)}
\end{align}

FOC:
\[\begin{gathered}l^*=\left[\left(\frac{1-\alpha\eta}{\beta}\right)\theta^{\frac{\alpha\eta}{\alpha\eta-1}}k^{\frac{\alpha}{\alpha\eta-1}}w\right]^{\frac{1-\alpha\eta}{\beta+\alpha\eta-1}}\\\pi(k,w)=\theta^{\frac{\alpha\eta}{1-\alpha\eta}}k^{\frac{\alpha}{1-\alpha\eta}}(l^{*})^{\frac{\beta}{1-\alpha\eta}}-wl^{*}\end{gathered}\]

That is to say:

\[\begin{gathered}\pi(k,w)=\left\{\theta^{\frac{\alpha\eta}{1-\alpha\eta}}\left[\left(\frac{1-\alpha\eta}{\beta}\right)\theta^{\frac{\alpha\eta}{\alpha\eta-1}}w\right]^{\frac{\beta}{(\beta+\alpha\eta-1)}}-w\left[\left(\frac{1-\alpha\eta}{\beta}\right)\theta^{\frac{\alpha\eta}{\alpha\eta-1}}w\right]^{\frac{1-\alpha\eta}{\beta+\alpha\eta-1}}\right\}k^{-\frac{\alpha}{\beta+\alpha\eta-1}}\\\pi(k,w)=\underbrace{\pi(w)k^{\frac{\alpha}{1-\beta-\alpha\eta}}}_{\text{\textit{accounting profit}}}\end{gathered}\]

Which:
\begin{align}
\pi(w)=\left\{\theta^{\frac{\alpha\eta}{1-\alpha\eta}}\left[\left(\frac{1-\alpha\eta}{\beta}\right)\theta^{\frac{\alpha\eta}{\alpha\eta-1}}w\right]^{\frac{\beta}{(\beta+\alpha\eta-1)}}-w\left[\left(\frac{1-\alpha\eta}{\beta}\right)\theta^{\frac{\alpha\eta}{\alpha\eta-1}}w\right]^{\frac{1-\alpha\eta}{\beta+\alpha\eta-1}}\right\}\label{(4)}
\end{align}

Further, the economic profit of the firm can be written as:
\[\pi(r,w)=\max_k\left\{\pi(w)k^{\frac{\alpha}{1-\beta-\alpha\eta}}-rk\right\}\]

Accordingly, the endogenous interest rate can be obtained:
\begin{align}
r=\frac{\alpha}{1-\beta-\alpha\eta}\pi(w)k^{\frac{\alpha+\beta+\alpha\eta-1}{1-\beta-\alpha\eta}}\label{(5)}
\end{align}

And:
\[\pi(w)=\left\{\theta^{\frac{\alpha\eta}{1-\alpha\eta}}\left[\left(\frac{1-\alpha\eta}{\beta}\right)\theta^{\frac{\alpha\eta}{\alpha\eta-1}}w\right]^{\frac{\beta}{(\beta+\alpha\eta-1)}}-w\left[\left(\frac{1-\alpha\eta}{\beta}\right)\theta^{\frac{\alpha\eta}{\alpha\eta-1}}w\right]^{\frac{1-\alpha\eta}{\beta+\alpha\eta-1}}\right\}\]

\noindent\textbf{\textit{Maximizing Capital Stock}}. The investment problem of the firm is as follows:
\begin{align}
\max_{\{i(t),k(t)\}}\int_0^\infty e^{-rt}\left[\pi(w(t))k(t)^{\frac{\alpha}{1-\beta-\alpha\eta}}-i(t)-\frac{1}{2}a\left(\frac{i(t)}{k(t)}-\delta\right)^2k(t)\right]dt\label{(6)}
\end{align}

s.t.:
\[\dot{k}(t)=i(t)-\delta k(t)\]
\[\pi(w(t))=\left\{\theta^{\frac{\alpha\eta}{1-\alpha\eta}}\left[\left(\frac{1-\alpha\eta}{\beta}\right)\theta^{\frac{\alpha\eta}{\alpha\eta-1}}w(t)\right]^{\frac{\beta}{\beta+\alpha\eta-1)}}-w(t)\left[\left(\frac{1-\alpha\eta}{\beta}\right)\theta^{\frac{\alpha\eta}{\alpha\eta-1}}w(t)\right]^{\frac{1-\alpha\eta}{\beta+\alpha\eta-1}}\right\}\]

The current-value Hamiltonian is written as:
\begin{align}
H=\underbrace{\pi(w(t))k(t)^{\frac{\alpha}{1-\beta-\alpha\eta}}}_{\text{\textit{accounting profit}}}-\underbrace{i(t)}_{investment}-\underbrace{\frac{1}{2}a\left[\frac{i(t)}{k(t)}-\delta\right]^2k(t)}_{\text{\textit{adjustment cost}}}+\underbrace{q(t)}_{multiplier}\underbrace{(i(t)-\delta k(t))}_{\dot{k}(t)}\label{(7)}
\end{align}

FOC:
\[\begin{gathered}\frac{\partial H}{\partial i}=0\\\frac{\partial H}{\partial k}=0\end{gathered}\]

Hence, according to \textit{Pontryagin's} maximum principle:

\[\begin{gathered}
i^*(t)=\arg\max_iH(k,i,q)\\k^*(t)=\arg\max_kH(k,i,q)
\end{gathered}\]

We can calculate and get:
\[\begin{gathered}1+a\left[\frac{i(t)}{k(t)}-\delta\right]=q(t)\\\frac{i(t)}{k(t)}=\delta+\frac{1}{a}[q(t)-1]\\\frac{\partial H}{\partial k}=\pi(w(t))\left(\frac{\alpha}{1-\beta-\alpha\eta}\right)k(t)^{\frac{\alpha+\beta+\alpha\eta-1}{1-\beta-\alpha\eta}}+\frac{1}{2}a\left[\left(\frac{i(t)}{k(t)}\right)^2-\delta^2\right]-q(t)\delta\\\dot{q}(t)=\frac{dq(t)}{dt}=(r+\delta)q(t)-\pi(w(t))\left(\frac{\alpha}{1-\beta-\alpha\eta}\right)k(t)^{\frac{\alpha+\beta+\alpha\eta-1}{1-\beta-\alpha\eta}}-\frac{1}{2}a\left[\left(\frac{i(t)}{k(t)}\right)^2-\delta^2\right]\end{gathered}\]

Given:
\[\left(\frac{i(t)}{k(t)}\right)^2=\delta^2+\frac{1}{a^2}[q(t)-1]^2+\frac{2\delta}{a}[q(t)-1]\]

Hence:
\[\dot{q}(t)=(r+\delta)q(t)-\pi(w(t))\left(\frac{\alpha}{1-\beta-\alpha\eta}\right)k(t)^{\frac{\alpha+\beta+\alpha\eta-1}{1-\beta-\alpha\eta}}-\frac{1}{2}a\left\{\frac{1}{a^2}[q(t)-1]^2+\frac{2\delta}{a}[q(t)-1]\right\}\]

The first-order condition for investment can be written as:
\[\underbrace{\dot{q}(t)=0}_{\text{\textit{steady-state investment}}}\]

Under steady-state conditions:
\begin{align}
k=\left\{\frac{(r+\delta)q-\frac{1}{2}a\left\{\frac{1}{a^2}[q(t)-1]^2+\frac{2\delta}{\alpha}[q(t)-1]\right\}}{\pi(w)\left(\frac{\alpha}{1-\beta-\alpha\eta}\right)}\right\}^{\frac{1-\beta-\alpha\eta}{\alpha+\beta+\alpha\eta-1}}\label{(8)}
\end{align}

Plugging $i(t)=\left\{\delta+\frac{1}{a}[q(t)-1]\right\}k(t)$:
\[\begin{gathered}\dot{k}(t)=\left\{\delta+\frac{1}{a}[q(t)-1]\right\}k(t)-\delta k(t)\\\frac{\dot{k}(t)}{k(t)}=\delta+\frac{1}{a}[q(t)-1]-\delta=\frac{1}{a}[q(t)-1]\end{gathered}\]

Then, we substitute the steady-state conditions $\dot{k}(t)=0$ and $q=1$ in this section:
\[k^*=\left(\frac{(r+\delta)(1-\beta-\alpha\eta)}{\alpha\left(\theta^{\frac{\alpha\eta}{1-\alpha\eta}}\left[\left(\frac{1-\alpha\eta}{\beta}\right)\theta^{\frac{\alpha\eta}{\alpha\eta-1}}w\right]^{\frac{\beta}{(\beta+\alpha\eta-1)}}-w\left[\left(\frac{1-\alpha\eta}{\beta}\right)\theta^{\frac{\alpha\eta}{\alpha\eta-1}}w\right]^{\frac{1-\alpha\eta}{\beta+\alpha\eta-1}}\right)}\right)^{\frac{1-\beta-\alpha\eta}{\alpha+\beta+\alpha\eta-1}}\]

\subsection{The Household Problem}
\textbf{\textit{Maximizing Utility}}. Assuming that the consumer utility function is a CRRA utility function, consumption is $c(t)$, the risk aversion coefficient $\sigma>1$, and the utility discount factor is $\rho$, the utility maximization problem is as follows:
\begin{align}
\max_{c,\dot{k}}\int_0^\infty e^{-\rho t}\left[\frac{c(t)^{1-\sigma}}{1-\sigma}\right]dt\label{(9)}
\end{align}

s.t.:
\[\begin{gathered}\dot{k}(t)=r(t)k(t)+w(t)l(t)-c(t)-\delta k(t)\\r(t)=\frac{\alpha}{1-\beta-\alpha\eta}\pi(w)k(t)^{\frac{\alpha+\beta+\alpha\eta-1}{1-\beta-\alpha\eta}}\\c(t)\geqslant0\end{gathered}\]

According to equation $(5)$, $r(t)$ is the endogenous interest rate and comes from the firm's investment problem. For the household, the current-value Hamiltonian is written as:
\begin{align}
H=\frac{c(t)^{1-\sigma}}{1-\sigma}+\lambda(t)[r(t)k(t)+w(t)l(t)-c(t)-\delta k(t)]\label{(10)}
\end{align}

FOC:
\[\frac{\partial H}{\partial c}=0\]

Hence, according to \textit{Pontryagin's} maximum principle:
\[c^*(t)=\arg\max_cH(k,c,\lambda)\]

And:
\[\begin{gathered}c(t)^{-\sigma}=\lambda(t)\\\dot{\lambda}(t)=\rho\lambda(t)-\frac{\partial H}{\partial k}\\\frac{\dot{\lambda}(t)}{\lambda(t)}=\rho-r(t)+\delta\\\frac{\dot{c}(t)}{c(t)}=\frac{r(t)-\rho-\delta}{\sigma}\end{gathered}\]

Plugging $r(t)$, we get:
\begin{align}
\frac{\dot{c}(t)}{c(t)}=\frac{\frac{\alpha}{1-\beta-\alpha\eta}\pi(w)k(t)^\frac{\alpha+\beta+\alpha\eta-1}{1-\beta-\alpha\eta}-\rho-\delta}{\sigma}\label{(11)}
\end{align}

\subsection{The Analytical Solution}
According to equation $(3)$, firms choose labor $l^*$ to maximize profits, hence we have $l(t)=l^*$ under steady state, that is:
\[l^*=\frac{\beta y}{(1-\alpha\eta)w}\]

We substitute $l^*$ into the production function with data and get:

\[\begin{gathered}
y=\theta^{\frac{\alpha\eta}{1-\alpha\eta}}k^{\frac{\alpha}{1-\alpha\eta}}\left(\frac{\beta y}{(1-\alpha\eta)w}\right)^{\frac{\beta}{1-\alpha\eta}}\\y^{1-\frac{\beta}{1-\alpha\eta}}=\theta^{\frac{\alpha\eta}{1-\alpha\eta}}k^{\frac{\alpha}{1-\alpha\eta}}\left(\frac{\beta}{1-\alpha\eta}\right)^{\frac{\beta}{1-\alpha\eta}}w^{-\frac{\beta}{1-\alpha\eta}}\\y=\theta^{\frac{\alpha\eta}{1-\beta-\alpha\eta}}{\left[\left(\frac{1-\alpha\eta}{\beta}\right)w\right]}^{\frac{\beta}{\beta+\alpha\eta-1}}k^{\frac{\alpha}{1-\beta-\alpha\eta}}
\end{gathered}\]

Now, according to the capital stock dynamics in market clear:
\[\dot{k}(t)=y(t)-c(t)-\delta k(t)\]

We have:
\[\dot{k}(t)=\theta^{\frac{\alpha\eta}{1-\beta-\alpha\eta}}{\left[\left(\frac{1-\alpha\eta}{\beta}\right)w\right]}^{\frac{\beta}{\beta+\alpha\eta-1}}k^{\frac{\alpha}{1-\beta-\alpha\eta}}-c(t)-\delta k(t)\]

Hence, the general equilibrium could be written as:
\[\begin{gathered}\frac{\dot{c}(t)}{c(t)}=\frac{\frac{\alpha}{1-\beta-\alpha\eta}\pi(w)k(t)^{\frac{\alpha+\beta+\alpha\eta-1}{1-\beta-\alpha\eta}}-\rho-\delta}{\sigma}\\\frac{\dot{k}(t)}{k(t)}=\frac{\theta^{\frac{\alpha\eta}{1-\beta-\alpha\eta}}\left[\left(\frac{1-\alpha\eta}{\beta}\right)w\right]^\frac{\beta}{(\beta+\alpha\eta-1)}k(t)^{\frac{\alpha}{1-\beta-\alpha\eta}}-c(t)-\delta k(t)}{k(t)}\\\pi(w)=\left\{\theta^{\frac{\alpha\eta}{1-\alpha\eta}}\left[\left(\frac{1-\alpha\eta}{\beta}\right)\theta^{\frac{\alpha\eta}{\alpha\eta-1}}w\right]^{\frac{\beta}{(\beta+\alpha\eta-1)}}-w\left[\left(\frac{1-\alpha\eta}{\beta}\right)\theta^{\frac{\alpha\eta}{\alpha\eta-1}}w\right]^{\frac{1-\alpha\eta}{\beta+\alpha\eta-1}}\right\}\end{gathered}\]

When this economic system reaches steady state, $\dot{c}(t)=0$ and $\dot{k}(t)=0$. Therefore, we can get the analytical solutions for equilibrium consumption $c^*$ and equilibrium capital $k^*$:
\[\begin{gathered}k^*=\left({\frac{(\rho+\delta)(1-\beta-\alpha\eta)}{\alpha\left(\theta^{\frac{\alpha\eta}{1-\alpha\eta}}\left[\left(\frac{1-\alpha\eta}{\beta}\right)\theta^{\frac{\alpha\eta}{\alpha\eta-1}}w\right]^{\frac{\beta}{(\beta+\alpha\eta-1)}}-w\left[\left(\frac{1-\alpha\eta}{\beta}\right)\theta^{\frac{\alpha\eta}{\alpha\eta-1}}w\right]^{\frac{1-\alpha\eta}{\beta+\alpha\eta-1}}\right)}}\right)^{\frac{1-\beta-\alpha\eta}{\alpha+\beta+\alpha\eta-1}}\\c^*=\theta^{\frac{\alpha\eta}{1-\beta-\alpha\eta}}\left[(\frac{1-\alpha\eta}{\beta})w\right]^{\frac{\beta}{(\beta+\alpha\eta-1)}}(k^*)^{\frac{\alpha}{1-\beta-\alpha\eta}}-\delta k^*\end{gathered}\]

In the general equilibrium analysis of this paper, given $r=r(t)$, which is the endogenous interest rate given by the firm's profit problem, the consumer solves the utility maximization problem and the firm solves the investment maximization problem according to $r(t)$.

\subsection{Numerical Simulation and Phase Diagram Analysis}
\textbf{\textit{Parameter Setting}}. According to the analytical solution $(c^*,k^*)$, the parameters of numerical simulation in this section are set as follows: We assume $0<\alpha+\beta\leqslant1$. Referring to the study of \citet{jones2020nonrivalry}, the output elasticity of data factors in the production function is distributed between $0.03-0.12$. Therefore, we set the capital output elasticity $\alpha = 0.75$ and the labor output elasticity $\beta = 0.2$; $w$ is a parameter measuring the wage level, $w = 1$; $\delta$ is the capital depreciation rate, $\delta = 0.08$; $\rho$ is the utility discount rate, $\rho = 0.07$. Meanwhile, for each steady state, $\theta\in(0,1)$ and $\eta\in(0,1)$.

The simulation results are shown in Figures 2 and Figure 3:

\begin{figure}[htbp]
\centering
\includegraphics[width=15cm]{B2.jpg}
\caption{\label{fig:B2}The Transition Dynamics of $k^*$ and $\eta(\theta)$ with Stationary $k^*$}
\end{figure}
\begin{figure}[htbp]
\centering
\includegraphics[width=15cm]{B3.jpg}
\caption{\label{fig:B3}The Transition Dynamics of $c^*$ and $\eta(\theta)$ with Stationary $c^*$}
\end{figure}

Figure 2 illustrates the dynamic transitions of equilibrium capital $k^*$ as the dataization rate $\theta$ and the technological progress $\eta$ evolve, along with the nexus between dataization and technological progress under a stationary capital stock, denoted by $\eta(\theta,k^*_i)$, where $k^*_i \in k^*$. We assume that $\eta(\theta)$ exists in the stationary state $\mathbb{K}^*$, where $\mathbb{K}^* \subset k^*, \sup \mathbb{K}^* = k^*_h, \inf \mathbb{K}^* = k^*_l$, and $k^*_h > k^*_l$. We find that:

(\textbf{a}) If we consider any specific level of $\theta_0 \in (0,1)$, an increase in $\eta$ will raise the equilibrium capital stock $k^*$, and this effect exhibits increasing marginal returns. However, if we consider any specific level of $\eta_0 \in (0,1)$, an increase in $\theta$ will not contribute to raising the equilibrium capital stock $k^*$. Actually, $\theta$ acts as a negative moderating factor on the effect of $\eta$ in raising $k^*$. Therefore, equilibrium capital stock $k^*$ can only maintain stable growth if we encourage both dataization activities and technological progress, that is, we should increase both $\theta$ and $\eta$ simultaneously.

(\textbf{b}) For $\eta(\theta, k^*_i<k^*_l)$ or $\eta(\theta, k^*_i>k^*_h)$, we did not find a significant correlation between dataization and technological progress. However, for $\eta(\theta, k^*_i\in\mathbb{K}^*)$, we find a clear positive correlation between dataization and technological progress, where an increase in $\theta$ significantly boosts $\eta$. This suggests that when the capital stock in the macroeconomy of our paper is stationary, increasing the proportion of dataization can effectively promote technological progress at a high level.

Figure 3 illustrates the dynamic transitions of equilibrium consumption $c^*$ as the dataization rate $\theta$ and technological progress $\eta$ evolve, along with the nexus between dataization and technological progress under stationary consumption, denoted by $\eta(\theta,c^*_i)$, where $c^*_i \in c^*$. We assume that $\eta(\theta)$ exists in the stationary state $\mathbb{C}^*$, where $\mathbb{C}^* \subset c^*, \sup \mathbb{C}^* = c^*_h, \inf \mathbb{C}^* = c^*_l$, and $c^*_h > c^*_l$. We find that:

(\textbf{a}) If we fix a specific level of $\theta_0 \in (0,1)$, there is an inverted U-shaped relationship between $\eta$ and equilibrium consumption $c^*$. This indicates that there is a threshold $\eta^*$ for the growth of equilibrium consumption driven by technological progress. Specifically, when $\eta>\eta^*$, encouraging technological progress will be detrimental to the growth of equilibrium consumption. Meanwhile, if we fix $\eta_0 \in (0,1)$ at any specific level, an increase in $\theta$ will hinder the growth of equilibrium consumption $c^*$. This is similar to the situation in Figure 2, $\theta$ exerts a negative moderating effect on the impact of $\eta$ on $c^*$. Therefore, we can conclude that $\theta$ exerts a negative moderating effect on the transitions in the general equilibrium model driven by $\eta$. Only a simultaneous increase in both $\theta$ and $\eta$ can ensure that the general equilibrium shifts stably in a positive direction to increase both equilibrium capital stock $k^*$ and equilibrium consumption $c^*$.

(\textbf{b}) For $\eta(\theta, c^*_i)$, the results are similar to those in Figure 1, when stationary consumption is at a normal level with $\eta(\theta, c^*_i\in\mathbb{C}^*)$, the nexus between dataization and technological progress follows a trend similar to an inverted U-shaped, and its threshold depends solely on the level of technological progress. Specifically, when technological progress is at a relatively low level, increasing the proportion of dataization can effectively promote technological progress. However, when technological progress is at a relatively high level, increasing the proportion of dataization fails to further promote technological progress. Meanwhile, this also explains the origin of the threshold observed in Figure 1. Since dataization could promote technological progress at low levels, the technological progress $\eta(\theta)$ in Figure 1 initially rises with increasing $\theta$ when it starts at a low level. However, once $\eta(\theta)$ continues to increase and enters the advanced level range, dataization begins to hinder technological progress at high levels. This results in $\eta(\theta)$ decreasing when $\theta$ increases.

\noindent\textbf{\textit{Phase Diagram}}. The two graphs in Figure 4 provide a phase diagram analysis of our study. Figure 4-a illustrates the shock from technological progress to the general equilibrium under a constant dataization. Figure 4-b illustrates the shock from dataization on general equilibrium under a constant technological progress.

In Figure 4-a, $E_1$ is the initial general equilibrium state. When technological progress increases before reaching its threshold $\eta^*$, $E_1$ shifts to $E_2$, leading to an increase in both the equilibrium capital stock and equilibrium consumption. When technological progress increases beyond its threshold $\eta^*$, $E_1$ shifts to $E_3$, where the equilibrium capital stock increases, but the equilibrium consumption decreases.

In Figure 4-b, $E_1$ is the initial general equilibrium state. Since dataization has a negative moderating effect on technological progress, as the proportion of dataization increases, $E_1$ shifts to $E_2$, leading to a reduction in both the equilibrium capital stock and equilibrium consumption.

\begin{figure}[htbp]
\centering
\includegraphics[width=15.5cm]{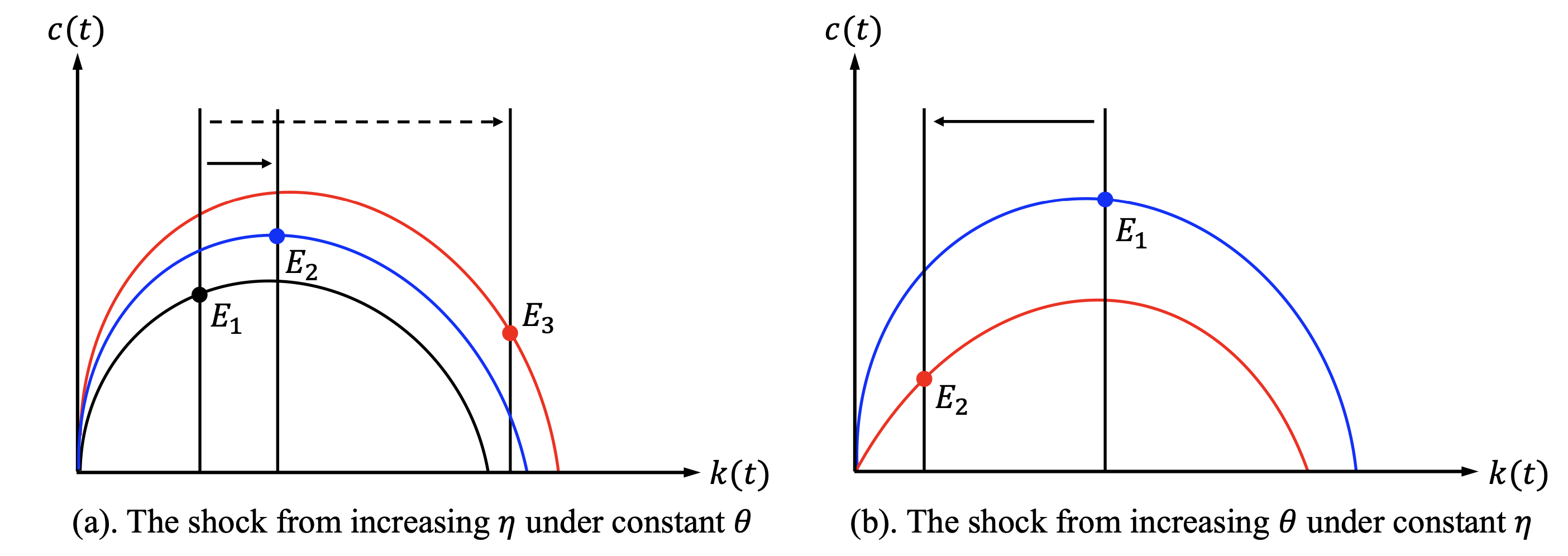}
\caption{\label{fig:B4}The Phase Diagram Analysis of General Equilibrium}
\end{figure}
\vspace{-0.3cm}

\section{Research Design and Approach}
\subsection{Econometric Hypothesis}
In this section, we will propose several hypotheses for empirical research based on the theoretical model discussed in this paper. Then, we will apply macroeconomic data of China from 2000 to 2021 to verify these hypotheses.

Based on the results from Section 3.2 of this paper, we find that technological progress can promote the growth of the equilibrium capital stock, and its relationship with equilibrium consumption follows an inverted U-shaped pattern. Furthermore, dataization has a negative moderating effect on the impact of technological progress on the transition of general equilibrium. Only when policies simultaneously encourage both dataization and technological progress can higher dataization proportions promote the growth of the equilibrium capital stock and equilibrium consumption. Therefore, we propose the following Hypothesis 1, Hypothesis 2 and Hypothesis 3:

\begin{Hypothesis}
Policy can enhance the social capital stock by promoting technological progress.
\end{Hypothesis}

\begin{Hypothesis}
The effect of a policy on boosting social consumption by promoting technological progress is limited. In economies with low levels of technological progress, it can significantly increase social consumption. However, in economies with high levels of technological progress, its effect on boosting social consumption will be weak.
\end{Hypothesis}

\begin{Hypothesis}
Policy could ensure stable growth in social capital stock and consumption only when it simultaneously encourages dataization and technological progress. Therefore, if a policy has already promoted the growth of social capital stock and consumption by increasing the proportion of dataization, it has also enhanced technological progress.
\end{Hypothesis}

Regarding the nexus between dataization and technological progress, the analysis in Section 2.2 preliminarily reveals and explains the existence of a threshold for dataization to enhance technological progress. Section 3.4 analyzes this issue in a general equilibrium framework. We find that, in a general equilibrium model with stationary capital stock, an increase in the dataization proportion can promote technological progress with increasing marginal returns. However, in a general equilibrium model with stationary consumption, an increase in the dataization proportion can promote technological progress at low levels while reducing it at high levels. Therefore, we further propose Hypothesis 4 and Hypothesis 5:

\begin{Hypothesis}
Policy can promote technological progress by increasing the proportion of dataization in stationary social capital stock.
\end{Hypothesis}

\begin{Hypothesis}
The effect of policy on promoting technological progress by increasing dataization proportion in stationary social consumption is limited. In the initial period of policy implementation, increasing the dataization proportion can effectively promote technological progress. In the final period of policy implementation, further increasing the dataization proportion fails to promote technological progress.
\end{Hypothesis}

\subsection{Variable and Data}
\textbf{\textit{Main Research Variables}}. We will conduct the empirical analysis using macroeconomic data from 216 prefecture-level cities and 4 municipalities directly under the central government in China from 2000 to 2021\footnote{We selected data from 2000 to 2021 for two main reasons: First, it ensures a complete and stable long panel data based on data availability. Second, it helps avoid the “AI shock.” Specifically, ChatGPT (GPT-3.5), a conversational AI model officially released by the American research institute OpenAI, went live on November 30, 2022. The emergence of this AI technology has significantly impacted the “dataization” and “technological progress” studied in our paper. Therefore, to avoid endogeneity issues caused by this impact, we will not use data from 2022 or later.}. We have four main research variables: Technological progress, dataization, social capital stock and consumption.

We will use two variables to represent technological progress. The first type of technological progress can be regarded as “inclusive technological progress”, for which we use the Digital Economy Development Index as the variable (\textbf{Digital}). The second type of technological progress can be regarded as “specialized technological progress”, for which we use the number of utility model patents granted related to the data economy (in tens of thousands) as the variable (\textbf{Tech}). Specifically:

For the measurement of $Digital$, we combine available data at the city level in China to assess this variable from two perspectives: Internet development and digital financial inclusion. First, we use internet penetration rate, employment in related sectors, related output and mobile phone penetration rate as indicators of internet development. Next, we use the “China Digital Inclusive Finance Index” as the metric for inclusive finance, this index is jointly compiled by the Center for Digital Finance at Peking University and Ant Financial Group. Finally, after standardizing the data for these five indicators and performing dimensionality reduction via principal component analysis, we calculate the comprehensive digital economy development index\footnote{The specific metrics corresponding to four indicators of the internet development are: The number of broadband internet subscribers per 100 people, the proportion of employees in the computer services and software industry relative to total urban employed personnel, total telecommunications services per capita, and the number of mobile phone subscribers per 100 people. The raw data for these indicators were obtained from the China City Statistical Yearbook.}.

For the measurement of $Tech$, we directly use the number of utility model patents granted in 220 Chinese cities between 2000 and 2021 that are related to the data economy (in units of $10^4$) as a variable to measure specialized technological progress. All of this patent data is sourced from the China National Intellectual Property Administration.

We use a policy related to data openness to represent dataization. We select the macroeconomic policy titled “Establishment of Regional Government Data Open Platforms”. We designate pilot cities for data openness as our experimental subjects, and the \textbf{DID} is constructed as the product of the policy’s time dummy variable and the treatment group dummy variable. The treatment group dummy variable is defined as follows: If a province or region implements the data openness policy, the value for that region is set to 1. Otherwise, it is set to 0. The time dummy variable is defined as follows: If a province or region implements the data openness policy in a certain year, the value for that region is set to 1 starting from that year, prior to that year, the value is set to 0.

Regarding the social capital stock and consumption, we use the prefecture level cities’ total fixed assets investment in $10^3$ billion of RMB (\textbf{Capital}). And the prefecture level cities’ total retail sales of consumer goods in $10^3$ billion of RMB (\textbf{Consumption}). Both of these two kinds of data come from the official websites of local governments and the China City Statistical Yearbook.

\noindent\textbf{\textit{Heterogeneous Groups}}. In macroeconomics, heterogeneity in income and consumption structures may affect the general economic equilibrium of a society. Therefore, besides identifying heterogeneity based on the level of technological progress, we will use provincial Gini coefficients and provincial Engel coefficients to identify heterogeneity in income and consumption structures within the general equilibrium, respectively. Specifically:

Regarding the heterogeneity of technological progress: We classify regions with $Digital\leqslant0.28$ and $Tech\leqslant0.005$ as low level groups, i.e., $\mathrm{\textbf{Low-type}}$. Conversely, we classify regions with $Digital>0.28$ and $Tech>0.005$ as high level groups, i.e., $\mathrm{\textbf{High-type}}$.

Regarding the heterogeneity in income and consumption structures: We directly use China’s provincial Gini coefficients to identify heterogeneity in household income structures, denoted as the \textbf{Gini Group}.

We identify general provinces with $Gini<0.4$ and the four provincial-level cities with $Gini<0.35$ as the low Gini coefficient group, they are denoted as $\mathrm{\textbf{G}(-)}$ with $Gini\ Group=0$. This indicates that the cities in these provinces belong to the group with relatively low income inequality. Similarly, we identify provinces with $Gini>0.4$ and provincial cities with $Gini>0.35$ as the high Gini coefficient group, they are denoted as $\mathrm{\textbf{G}(+)}$ with $Gini\ Group=1$. This indicates that cities in these provinces belong to the group with relatively high income inequality\footnote{The Gini coefficient is a variable that measures the equality of income distribution among residents in a country or region, ranging from 0 to 1. A value of 0 represents absolute equality, where everyone has exactly the same income. 1 represents absolute inequality, where all income is concentrated in the hands of a single individual. In reality, this variable typically fluctuates between 0.2 and 0.6. In our paper, the Gini coefficient is calculated by sorting night-time light values and then computing the difference in the area under the Lorenz curve between cumulative population and light intensity. Using night-time light data to calculate the Gini coefficient is a common method that provides a perspective on income distribution disparities. Our night-time light data primarily comes from satellite observations (DMSP-OLS $\&$ NPP-VIIRS) of the Earth at night. These satellites capture the brightness of lights on the ground to reflect the intensity and distribution of human activity.}.

We use the deviation from 1 in the ratio of the Engel coefficient for urban households ($\%$) to that for rural households ($\%$) at the provincial level in China to identify heterogeneity in household consumption structures, denoted as the \textbf{Engel Group}\footnote{The Engel coefficient is used to measure the standard of living in a country or region, it refers to the proportion of food expenditure in total household consumption expenditure. The data on the Engel coefficient used in our paper are all sourced from China’s Provincial Statistical Yearbooks and Provincial Statistical Bulletins.}.

We first calculate the ratio of the Engel coefficient for urban households ($\%$) to that for rural households ($\%$) at the provincial level in China. We then subtract 1 from this ratio and take its absolute value, defining it as a variable to identify heterogeneity in consumption structures. We classify cities in provinces where the $Engel<0.1$ as belonging to the group with low consumption structure inequality, they are denoted as $\mathrm{\textbf{E}(-)}$ with $Engel\ Group=0$. We classify cities in provinces where the $Engel>0.1$ as belonging to the group with high consumption structure inequality, they are denoted as $\mathrm{\textbf{E}(+)}$ with $Engel\ Group=1$.

\noindent\textit{\textbf{Control Variables}}. We denote the control variables group as $\mathbb{C}$. The measurements of control variables come from the Official Website of Regional Government and the China City Statistical Yearbook. We select gross domestic product / $10^4$ RMB (\textbf{GDP}), regional household population / $10^4$ people (\textbf{Population}), the number of industrial enterprises above scale / $1$ enterprise (\textbf{Company}), total wages of employees on board / $10^4$ RMB (\textbf{Salary}), total foreign capital utilization / $10^4$ RMB (\textbf{Foreign}), total telecommunication services / $10^4$ RMB (\textbf{Teleservice}) as the control variables. And all control variables are processed by logarithmic transformation\footnote{$***$, $**$ and $*$ indicate significance at the $1\%$, $5\%$ and $10\%$ levels, respectively. Standard errors are shown in parentheses. Please refer to \hyperref[Appendix A]{Appendix A} for the complete econometric table with control variables.}.

\subsection{Econometric Model}
\textit{\textbf{Benchmark Regression Model}}. We will use the benchmark regression model with two-way fixed effects (City $\&$ Year) to analyze the effects of technological progress $Digital_{ijt}$ and $Tech_{ijt}$ on social capital stock ($Capital_{ijt}$) and social consumption ($Consumption_{ijt}$), respectively, where $j\in\{\mathrm{\textbf{Low-type}}, \mathrm{\textbf{High-type}}\}$. $\sum_{i}\mathbb{P}$ is the city fixed effect. $\sum_{t}\mathbb{Y}$ is the time fixed effect.
\[\begin{gathered}Capital_{ijt}=\alpha_0+\beta_0Digital_{ijt}+\sum_{k_0}\gamma_{k_0}\mathbb{C}+\sum_i\mathbb{P}+\sum_t\mathbb{Y}+\varepsilon_{it}\\Consumption_{ijt}=\alpha_1+\beta_1Digital_{ijt}+\sum_{k_1}\gamma_{k_1}\mathbb{C}+\sum_i\mathbb{P}+\sum_t\mathbb{Y}+\varepsilon_{it}\end{gathered}\]

\[\begin{gathered}Capital_{ijt}=\alpha_2+\beta_2Tech_{ijt}+\sum_{k_2}\gamma_{k_2}\mathbb{C}+\sum_i\mathbb{P}+\sum_t\mathbb{Y}+\varepsilon_{it}\\Consumption_{ijt}=\alpha_3+\beta_3Tech_{ijt}+\sum_{k_3}\gamma_{k_3}\mathbb{C}+\sum_i\mathbb{P}+\sum_t\mathbb{Y}+\varepsilon_{it}\end{gathered}\]

\noindent\textbf{\textit{Difference-in-Difference Model}}. Based on the framework of the benchmark regression model, we employ a DID model to evaluate policy impacts on the economy and attempt to analyze the following issues: (a) The effects of dataization $DID_{ilt}$ on social capital stock ($Capital_{ilt}$) and social consumption ($Consumption_{ilt}$). (b) The impact of dataization $DID_{ilt}$ on technological progress $Digital_{ilt}$ and $Tech_{ilt}$ with stationary capital stock. (c) The impact of dataization $DID_{ilt}$ on technological progress $Digital_{ilt}$ and $Tech_{ilt}$ with stationary consumption. Here, $l\in\{\mathrm{\textbf{G}(+)},\mathrm{\textbf{G}(-)},\mathrm{\textbf{E}(+)},\mathrm{\textbf{E}(-)}\}$.
\[\begin{gathered}Capital_{ilt}=\alpha_4+\beta_4DID_{ilt}+\sum_{k_4}\gamma_{k_4}\mathbb{C}+\sum_i\mathbb{P}+\sum_t\mathbb{Y}+\varepsilon_{it}\\Consumption_{ilt}=\alpha_5+\beta_5DID_{ilt}+\sum_{k_5}\gamma_{k_5}\mathbb{C}+\sum_i\mathbb{P}+\sum_t\mathbb{Y}+\varepsilon_{it}\\Digital_{ilt}=\alpha_6+\beta_6DID_{ilt}+\mu_1Capital_{ilt}+\sum_{k_6}\gamma_{k_6}\mathbb{C}+\sum_i\mathbb{P}+\sum_t\mathbb{Y}+\varepsilon_{it}\\Digital_{ilt}=\alpha_7+\beta_7DID_{ilt}+\mu_2Consumption_{ilt}+\sum_{k_7}\gamma_{k_7}\mathbb{C}+\sum_i\mathbb{P}+\sum_t\mathbb{Y}+\varepsilon_{it}\\Tech_{ilt}=\alpha_8+\beta_8DID_{ilt}+\mu_3Capital_{ilt}+\sum_{k_8}\gamma_{k_8}\mathbb{C}+\sum_i\mathbb{P}+\sum_t\mathbb{Y}+\varepsilon_{it}\\Tech_{ilt}=\alpha_9+\beta_9DID_{ilt}+\mu_4Consumption_{ilt}+\sum_{k_9}\gamma_{k_9}\mathbb{C}+\sum_i\mathbb{P}+\sum_t\mathbb{Y}+\varepsilon_{it}\end{gathered}\]

\noindent\textbf{\textit{Moderating Effect Model}}. We use the moderating effects model to analyze the moderating effects of dataization $DID_{ijt}$ to the impacts from technological progress $Digital_{ijt}$ and $Tech_{ijt}$ on the social capital stock ($Capital_{ijt}$) and social consumption ($Consumption_{ijt}$), respectively, where $j\in\{\mathrm{\textbf{Low-type}},\mathrm{\textbf{High-type}}\}$, and the moderating variables are $\mathbb{X}=\{\mathbf{X_1}, \mathbf{X_2}\}$.
\[\begin{gathered}\mathbf{X_1}=DID_{ijt}\times Digital_{ijt}\\Capital_{ijt}=\alpha_{10}+\beta_{10}Digital_{ijt}+\lambda_1\mathbf{X_1}+\phi_1DID_{ijt}+\sum_{k_{10}}\gamma_{k_{10}}\mathbb{C}+\sum_i\mathbb{P}+\sum_t\mathbb{Y}+\varepsilon_{it}\\Consumption_{ijt}=\alpha_{11}+\beta_{11}Digital_{ijt}+\lambda_{2}\mathbf{X}_{1}+\phi_{2}DID_{ijt}+\sum_{k_{11}}\gamma_{k_{11}}\mathbb{C}+\sum_{i}\mathbb{P}+\sum_{t}\mathbb{Y}+\varepsilon_{it}\\\mathbf{X_2}=DID_{ijt}\times Tech_{ijt}\\Capital_{ijt}=\alpha_{12}+\beta_{12}Tech_{ijt}+\lambda_3\mathbf{X_2}+\phi_3DID_{ijt}+\sum_{k_{12}}\gamma_{k_{12}}\mathbb{C}+\sum_i\mathbb{P}+\sum_t\mathbb{Y}+\varepsilon_{it}\\Consumption_{ijt}=\alpha_{13}+\beta_{13}Tech_{ijt}+\lambda_{4}\mathbf{X}_{2}+\phi_{4}DID_{ijt}+\sum_{k_{13}}\gamma_{k_{13}}\mathbb{C}+\sum_{i}\mathbb{P}+\sum_{t}\mathbb{Y}+\varepsilon_{it}\end{gathered}\]

In the next section, we will first analyze the aggregate econometric effects of technological progress and dataization policy on the social capital stock and consumption, as well as the nexus between technological progress and dataization policy with stationary conditions. We will then study the heterogeneous econometric effects based on the heterogeneity in technological progress, income structure and consumption structure.

\section{Empirical Evidence and Analysis}
\subsection{Descriptive Analysis}
The four graphs in Figure 5 and Figure 6 provide a preliminary illustration showing the nonlinear effects of technological progress on capital stock and consumption, where the blue samples represent the state prior to the dataization policy shock and the red samples represent the state following the dataization policy shock.
\begin{figure}[htbp]
\centering
\includegraphics[width=15cm]{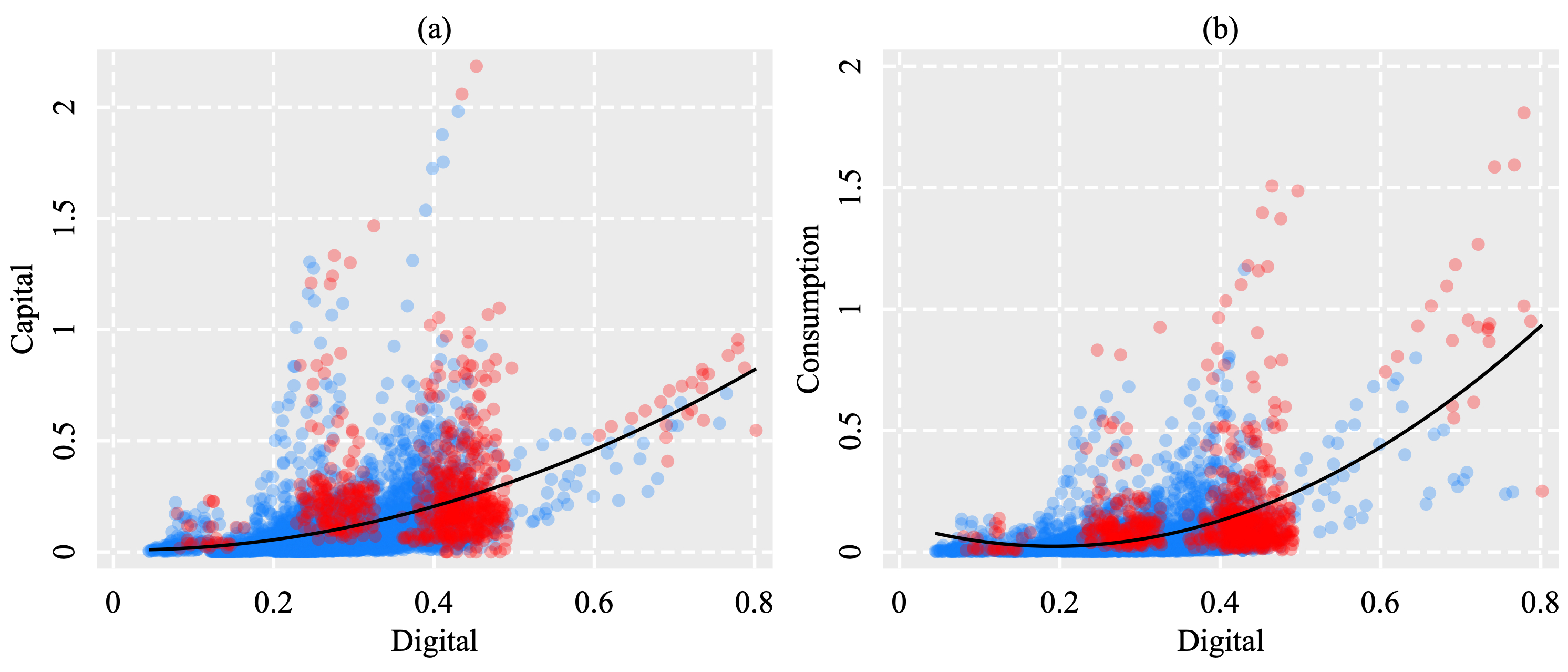}
\caption{\label{fig:A1}The Nonlinear Effects from Digital to Capital and Consumption}
\end{figure}
\vspace{-0.4cm}
\begin{figure}[htbp]
\centering
\includegraphics[width=15cm]{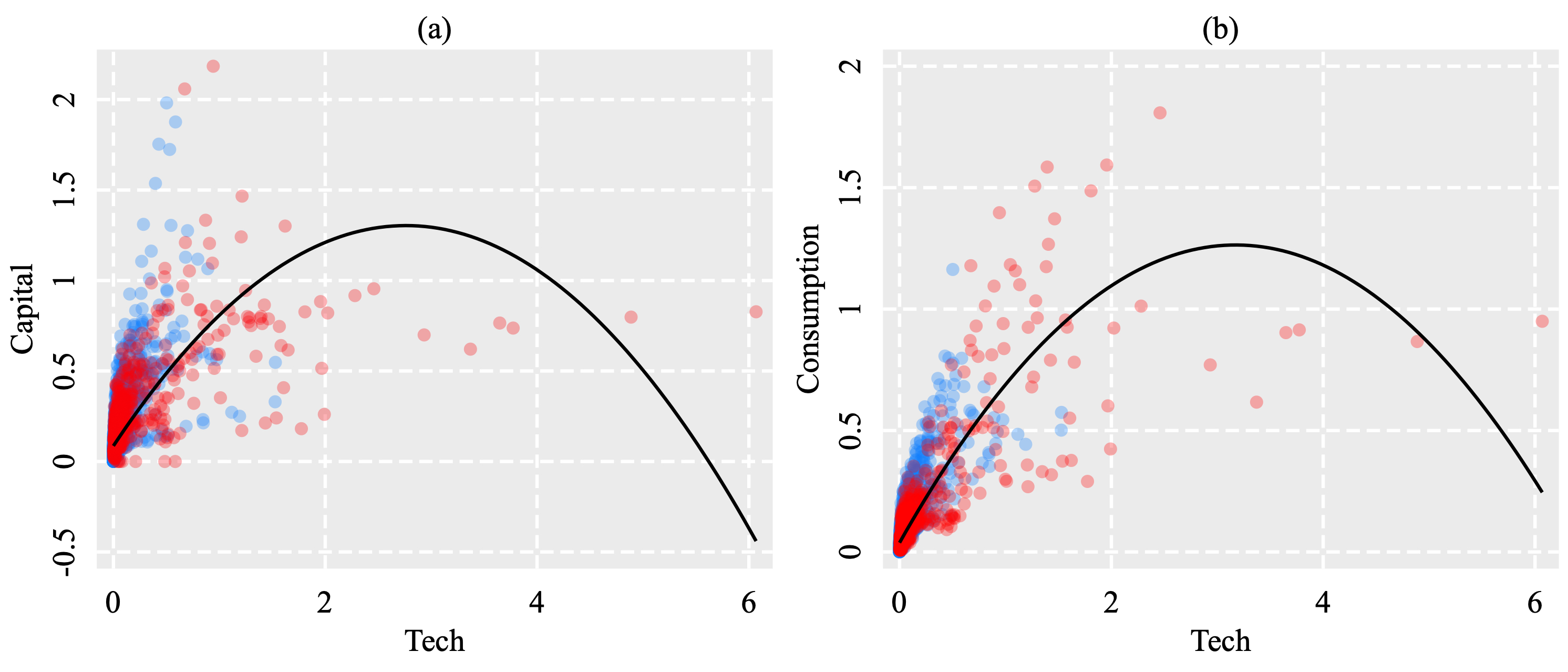}
\caption{\label{fig:A2}The Nonlinear Effects from Tech to Capital and Consumption}
\end{figure}

We find that inclusive technological progress $Digital$ promotes the growth of capital stock and consumption with increasing marginal returns, as shown in Figures 5-a and 5-b. However, the relationship between $Tech$ and capital stock or consumption is inverted U-shaped, as shown in Figures 6-a and 6-b, indicating that high level of specialized technological progress is detrimental to the growth of capital stock or consumption.

\subsection{Aggregate Effect Analysis}
Table 1, Table 2 and Table 3 respectively report the effects of technological progress on the social capital stock and consumption. And the effects of dataization policy shocks on the social capital stock and consumption.

We find that whether it is inclusive technological progress ($Digital$) or specialized technological progress ($Tech$), they can both significantly promote the growth of social capital stock ($Capital$), therefore, Hypothesis 1 is verified. At the same time, regarding the aggregate effect, both types of technological progress can significantly promote the growth of social consumption ($Consumption$), thus, part of Hypothesis 2 is supported. Furthermore, based on Table 3 and the dynamic effects of the policy in Figure 7, we can conclude that the implementation of dataization policy can significantly promote the growth of both social capital stock ($Capital$) and social consumption ($Consumption$) simultaneously. However, this growth effect exhibits a lagged period, that is, capital stock and consumption often do not show significant increases until several stages after policy implementation. This may be related to the fact that establishing databases to serve economic development requires time for construction and data cleaning.
\begin{table}[htbp]
\caption{The Aggregate Regression Effect of Digital on Capital and Consumption}
  \label{t1}
  \small
  \renewcommand{\arraystretch}{1.2}
  \centering
  \begin{tabularx}{0.9\textwidth}{l *{4}{>{\centering\arraybackslash}X}}
    \toprule
    & (1) & (2) & (3) & (4) \\
    & Capital & Capital & Consumption & Consumption \\
    \midrule
    Digital  & 1.251$^{***}$ & 0.407$^{***}$ & 0.871$^{***}$ & 0.632$^{***}$ \\
             & (0.037)       & (0.060)       & (0.027)       & (0.046)       \\
    Constant & -0.417$^{***}$& 0.147$^{**}$  & -0.114$^{***}$& 0.062         \\
             & (0.027)       & (0.054)       & (0.020)       & (0.041)       \\
    Controls & \checkmark    & \checkmark    & \checkmark    & \checkmark    \\
    City     & \checkmark    & \checkmark    & \checkmark    & \checkmark    \\
    Year     & $\times$      & \checkmark    & $\times$      & \checkmark    \\
    $R^2$    & 0.508         & 0.555         & 0.401         & 0.414         \\
    Number   & 4426          & 4426          & 4426          & 4426          \\
    \bottomrule
  \end{tabularx}
\end{table}
\begin{table}[htbp]
\caption{The Aggregate Regression Effect of Tech on Capital and Consumption}
  \label{t2}
  \small
  \renewcommand{\arraystretch}{1.2}
  \centering
  \begin{tabularx}{0.9\textwidth}{l *{4}{>{\centering\arraybackslash}X}}
    \toprule
    & (1) & (2) & (3) & (4) \\
    & Capital & Capital & Consumption & Consumption \\
    \midrule
    Tech     & 0.290$^{***}$ & 0.243$^{***}$ & 0.312$^{***}$ & 0.290$^{***}$ \\
             & (0.009)       & (0.008)       & (0.005)       & (0.005)       \\
    Constant & -0.621$^{***}$& 0.315$^{***}$ & -0.213$^{***}$& 0.298$^{***}$ \\
             & (0.025)       & (0.047)       & (0.015)       & (0.030)       \\
    Controls & \checkmark    & \checkmark    & \checkmark    & \checkmark    \\
    City     & \checkmark    & \checkmark    & \checkmark    & \checkmark    \\
    Year     & $\times$      & \checkmark    & $\times$      & \checkmark    \\
    $R^2$    & 0.504         & 0.637         & 0.596         & 0.668         \\
    Number   & 4426          & 4426          & 4426          & 4426          \\
    \bottomrule
  \end{tabularx}
\end{table}

\begin{table}[htbp]
\caption{The Aggregate Regression Effect of DID on Capital and Consumption }
  \label{t3}
  \small
  \renewcommand{\arraystretch}{1.2}
  \centering
  \begin{tabularx}{0.9\textwidth}{l *{4}{>{\centering\arraybackslash}X}}
    \toprule
    & (1) & (2) & (3) & (4) \\
    & Capital & Capital & Consumption & Consumption \\
    \midrule
    DID      & 0.096$^{***}$ & 0.001          & 0.069$^{***}$ & 0.032$^{***}$ \\
             & (0.006)       & (0.007)        & (0.005)       & (0.006)       \\
    Constant & -0.667$^{***}$& 0.244$^{***}$  & -0.286$^{***}$& 0.205$^{***}$ \\
             & (0.028)       & (0.052)        & (0.020)       & (0.041)       \\
    Controls & \checkmark    & \checkmark     & \checkmark    & \checkmark    \\
    City     & \checkmark    & \checkmark     & \checkmark    & \checkmark    \\
    Year     & $\times$      & \checkmark     & $\times$      & \checkmark    \\
    $R^2$    & 0.406         & 0.550          & 0.292         & 0.393         \\
    Number   & 4426          & 4426           & 4426          & 4426          \\
    \bottomrule
  \end{tabularx}
\end{table}

\begin{figure}[htbp]
\centering
\includegraphics[width=15.5cm]{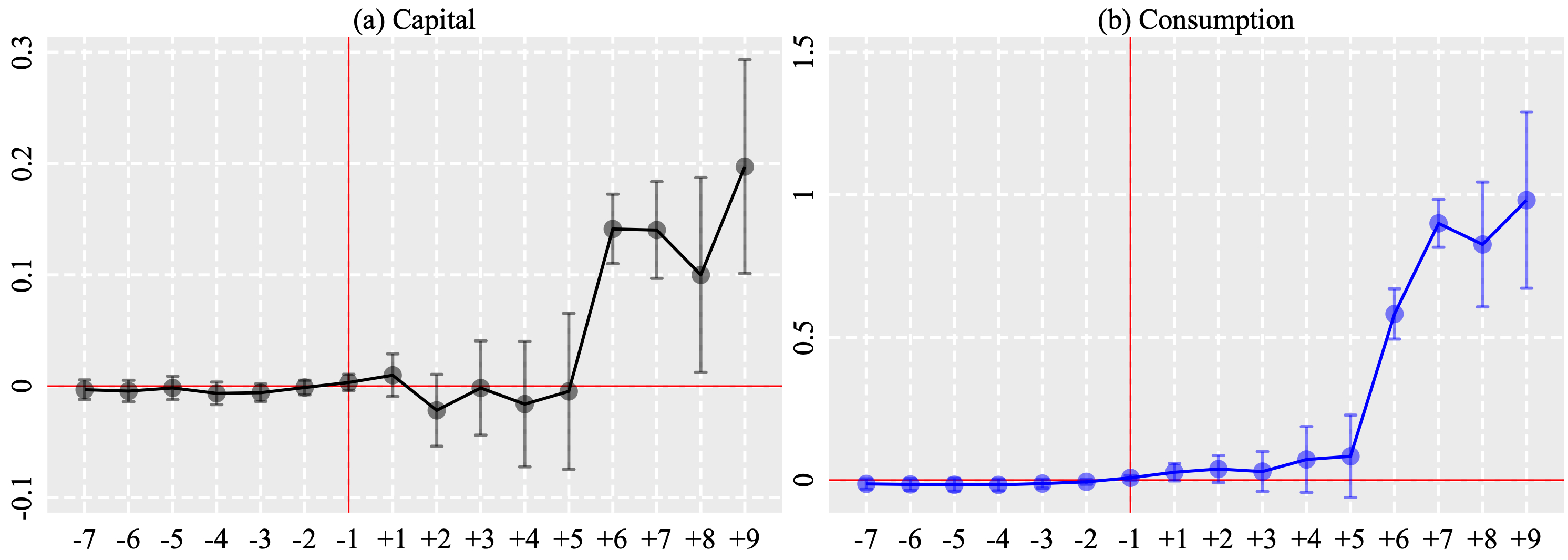}
\caption{\label{fig:A3}The Dynamic Effect of Dataization Policy on Capital and Consumption}
\end{figure}

Table 4 and Table 5 show the impact of dataization policy shocks on technological progress with stationary capital stock and consumption. Figure 8 and Figure 9 present the dynamic effects of the dataization policy.

According to Figure 8-a and Figure 8-b, we find that dataization policy can significantly promote technological progress with stationary capital stock, whether it is inclusive technological progress ($Digital$) or specialized technological progress ($Tech$).  

At the same time, based on Figure 9-a and Figure 9-b, we can see that the effect of dataization policy on promoting technological progress with stationary consumption is limited. In the early stages of policy implementation, the policy's role in promoting the growth of $Digital$ and $Tech$ is clear and significant. However, in the later stages of policy implementation, the policy's role in promoting the growth of $Digital$ and $Tech$ is not significant and may even turn into a negative effect. Therefore, we can preliminarily verify Hypothesis 4 and Hypothesis 5.

Furthermore, since dataization policy can significantly promote the growth of social capital stock and social consumption, and they can also promote technological progress to some extent. Hence, the conclusion regarding Hypothesis 3 is also reasonable.
\newpage
\begin{table}[htbp]
\caption{The Dataization Shock on Technological Progress with Stationary Capital }
  \label{t8}
  \small
  \renewcommand{\arraystretch}{1.2}
  \centering
  \begin{tabularx}{0.9\textwidth}{l *{4}{>{\centering\arraybackslash}X}}
    \toprule
    & (1) & (2) & (3) & (4) \\
    & Digital & Digital & Tech & Tech \\
    \midrule
    DID      & 0.041$^{***}$ & 0.003          & 0.031$^{***}$ & 0.085$^{***}$ \\
             & (0.002)       & (0.002)        & (0.009)       & (0.012)       \\
    Capital  & 0.150$^{***}$ & 0.026$^{***}$  & 0.703$^{***}$ & 0.797$^{***}$ \\
             & (0.005)       & (0.004)        & (0.022)       & (0.025)       \\
    Constant & -0.115$^{***}$& 0.233$^{***}$  & 0.150$^{***}$ & -0.510$^{***}$\\
             & (0.010)       & (0.013)        & (0.043)       & (0.085)       \\
    Controls & \checkmark    & \checkmark     & \checkmark    & \checkmark    \\
    City     & \checkmark    & \checkmark     & \checkmark    & \checkmark    \\
    Year     & $\times$      & \checkmark     & $\times$      & \checkmark    \\
    $R^2$    & 0.733         & 0.875          & 0.333         & 0.359         \\
    Number   & 4426          & 4426           & 4426          & 4426          \\
    \bottomrule
  \end{tabularx}
\end{table}
\begin{table}[htbp]
\caption{The Dataization Shock on Technological Progress with Stationary Consumption}
  \label{t9}
  \small
  \renewcommand{\arraystretch}{1.2}
  \centering
  \begin{tabularx}{0.9\textwidth}{l *{4}{>{\centering\arraybackslash}X}}
    \toprule
    & (1) & (2) & (3) & (4) \\
    & Digital & Digital & Tech & Tech \\
    \midrule
    DID          & 0.041$^{***}$ & 0.001          & -0.004         & 0.035$^{***}$ \\
                 & (0.002)       & (0.002)        & (0.008)        & (0.010)       \\
    Consumption  & 0.196$^{***}$ & 0.068$^{***}$  & 1.471$^{***}$  & 1.571$^{***}$ \\
                 & (0.007)       & (0.005)        & (0.025)        & (0.027)       \\
    Constant     & -0.159$^{***}$& 0.225$^{***}$  & 0.102$^{**}$   & -0.638$^{***}$\\
                 & (0.009)       & (0.013)        & (0.034)        & (0.070)       \\
    Controls     & \checkmark    & \checkmark     & \checkmark     & \checkmark    \\
    City         & \checkmark    & \checkmark     & \checkmark     & \checkmark    \\
    Year         & $\times$      & \checkmark     & $\times$       & \checkmark    \\
    $R^2$        & 0.728         & 0.879          & 0.542          & 0.565         \\
    Number       & 4426          & 4426           & 4426           & 4426          \\
    \bottomrule
  \end{tabularx}
\end{table}
\begin{figure}[ht!]
\centering
\includegraphics[width=15.5cm]{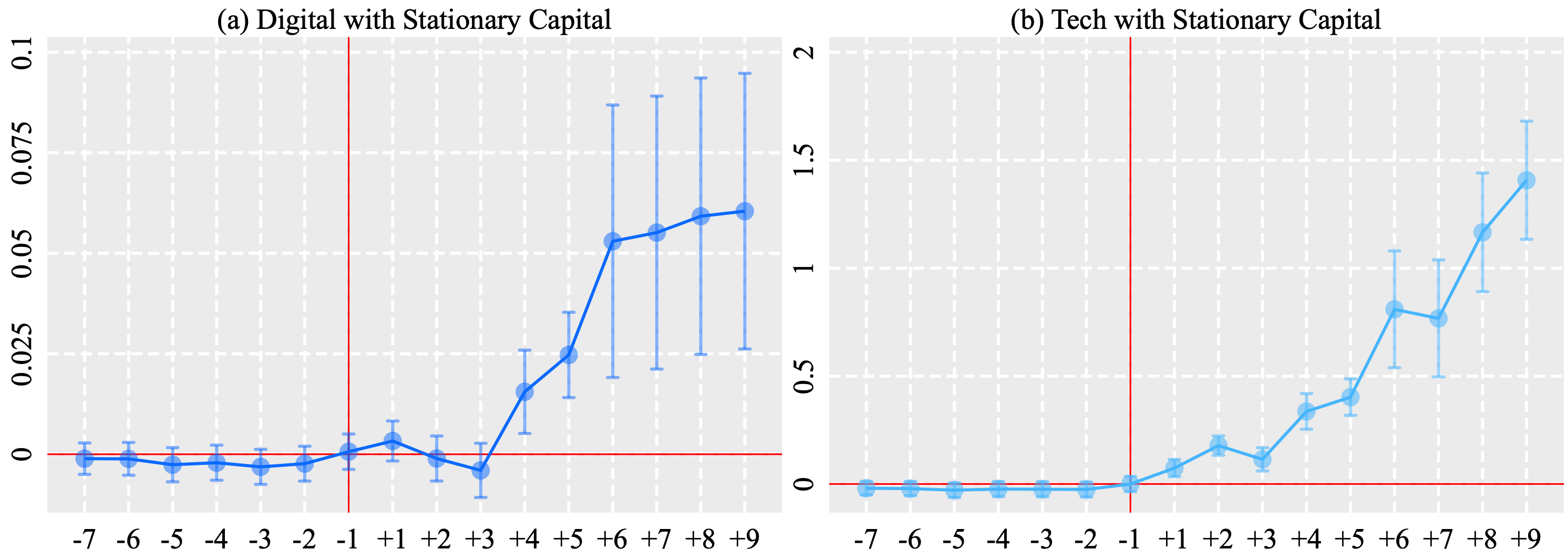}
\caption{\label{fig:A4}The Dynamics of Policy on Technological Progress with Stationary Capital}
\end{figure}

\begin{figure}[ht!]
\centering
\includegraphics[width=15.5cm]{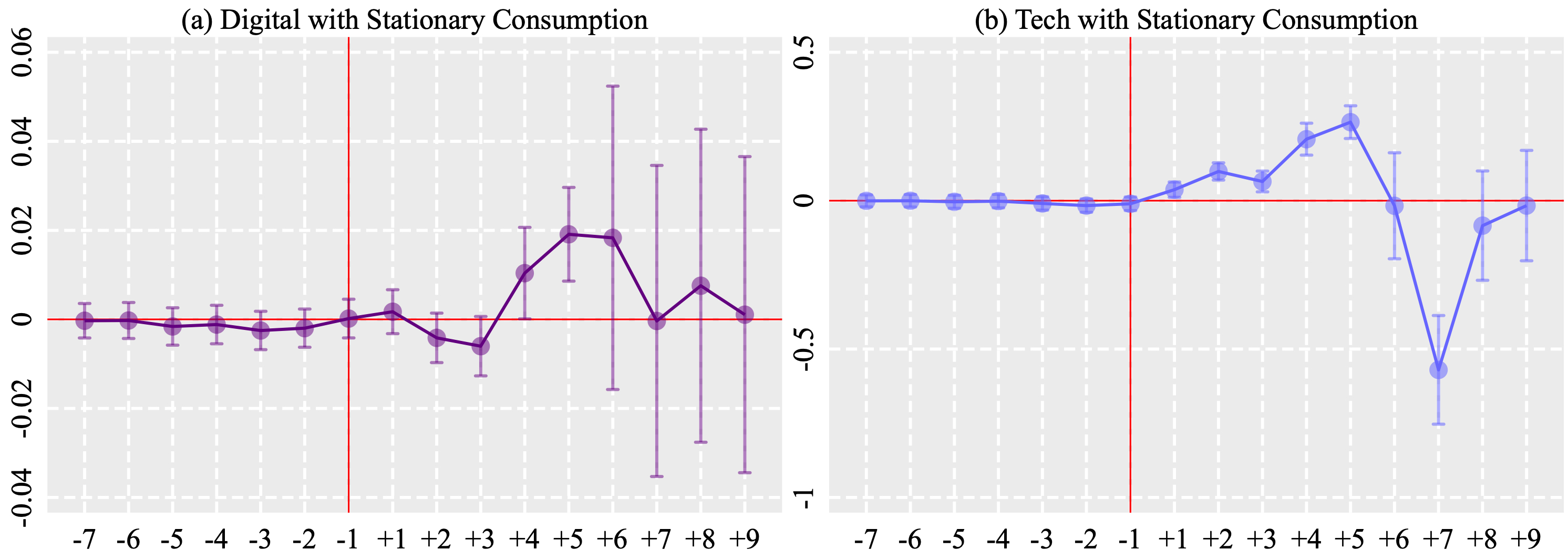}
\caption{\label{fig:A5}The Dynamics of Policy on Technological Progress with Stationary Consumption}
\end{figure}

\subsection{Moderating Effect Analysis}
Table 6, Table 7, Table 8 and Table 9 present the results of aggregate moderating effects from dataization policy. We find that:

For inclusive technological progress $Digital$, when time-fixed effects are not included, dataization policy has a significant negative moderating effect on the growth of social capital stock ($Capital$) promoted by $Digital$, which is consistent with the results in Figure 2. However, when time-fixed effects are included, dataization policy has a significant positive moderating effect on the growth of social capital stock ($Capital$) promoted by $Digital$. Furthermore, as shown in Table 7, dataization policy also exerts a significant positive moderating effect on the growth of social consumption ($Consumption$) promoted by $Digital$. This seems to be inconsistent with the previous conclusion. Therefore, we need to examine heterogeneity to reach further conclusions.

\begin{table}[htbp]
\caption{The Aggregate Moderating Effect from $\mathbf{X_1}$ on Capital}
  \label{t4}
  \small
  \renewcommand{\arraystretch}{1.2}
  \centering
  \begin{tabularx}{0.9\textwidth}{l *{4}{>{\centering\arraybackslash}X}}
    \toprule
    & (1) & (2) & (3) & (4) \\
    & Capital & Capital & Capital & Capital \\
    \midrule
    Digital  & 1.382$^{***}$ & 1.232$^{***}$ & 0.334$^{***}$ & 0.303$^{***}$ \\
             & (0.028)       & (0.041)       & (0.069)       & (0.069)       \\
    $\mathbf{X_1}$       & -0.162$^{***}$& -0.160$^{***}$& 0.170$^{***}$ & 0.136$^{**}$  \\
             & (0.040)       & (0.042)       & (0.043)       & (0.044)       \\
    DID      & 0.095$^{***}$ & 0.088$^{***}$ & -0.061$^{***}$& -0.053$^{**}$ \\
             & (0.015)       & (0.016)       & (0.018)       & (0.018)       \\
    Constant & -0.287$^{***}$& -0.405$^{***}$& -0.067$^{***}$& 0.174$^{**}$  \\
             & (0.008)       & (0.027)       & (0.015)       & (0.055)       \\
    Controls & $\times$      & \checkmark    & $\times$      & \checkmark    \\
    City     & \checkmark    & \checkmark    & \checkmark    & \checkmark    \\
    Year     & $\times$      & $\times$      & \checkmark    & \checkmark    \\
    $R^2$    & 0.505         & 0.513         & 0.549         & 0.556         \\
    Number   & 4426          & 4426          & 4426          & 4426          \\
    \bottomrule
  \end{tabularx}
\end{table}
\newpage
\begin{table}[htbp]
\caption{The Aggregate Moderating Effect from $\mathbf{X_1}$ on Consumption}
  \label{t5}
  \small
  \renewcommand{\arraystretch}{1.2}
  \centering
  \begin{tabularx}{0.9\textwidth}{l *{4}{>{\centering\arraybackslash}X}}
    \toprule
    & (1) & (2) & (3) & (4) \\
    & Consumption & Consumption & Consumption & Consumption \\
    \midrule
    Digital  & 0.700$^{***}$ & 0.712$^{***}$ & 0.317$^{***}$ & 0.267$^{***}$ \\
             & (0.020)       & (0.030)       & (0.052)       & (0.051)       \\
    $\mathbf{X_1}$       & 0.369$^{***}$ & 0.336$^{***}$ & 0.496$^{***}$ & 0.466$^{***}$ \\
             & (0.029)       & (0.030)       & (0.032)       & (0.033)       \\
    DID      & -0.105$^{***}$& -0.095$^{***}$& -0.153$^{***}$& -0.147$^{***}$\\
             & (0.011)       & (0.012)       & (0.013)       & (0.014)       \\
    Constant & -0.138$^{***}$& -0.126$^{***}$& -0.054$^{***}$& 0.146$^{***}$ \\
             & (0.006)       & (0.019)       & (0.011)       & (0.041)       \\
    Controls & $\times$      & \checkmark    & $\times$      & \checkmark    \\
    City     & \checkmark    & \checkmark    & \checkmark    & \checkmark    \\
    Year     & $\times$      & $\times$      & \checkmark    & \checkmark    \\
    $R^2$    & 0.413         & 0.423         & 0.430         & 0.445         \\
    Number   & 4426          & 4426          & 4426          & 4426          \\
    \bottomrule
  \end{tabularx}
\end{table}

For specialized technological progress $Tech$, we find that dataization policy has a significant negative moderating effect on the growth of social capital stock ($Capital$) and social consumption ($Consumption$) promoted by $Tech$ . This suggests that, for this type of technological progress, dataization policy weakens the impact of technological progress on capital stock and consumption, which is basically consistent with the results in Figure 2 and Figure 3. Similarly, we also need to examine the heterogeneity of this result.
\begin{table}[htbp]
\caption{The Aggregate Moderating Effect from $\mathbf{X_2}$ on Capital}
  \label{t6}
  \small
  \renewcommand{\arraystretch}{1.2}
  \centering
  \begin{tabularx}{0.9\textwidth}{l *{4}{>{\centering\arraybackslash}X}}
    \toprule
    & (1) & (2) & (3) & (4) \\
    & Capital & Capital & Capital & Capital \\
    \midrule
    Tech     & 1.039$^{***}$ & 0.840$^{***}$ & 0.656$^{***}$ & 0.665$^{***}$ \\
             & (0.023)       & (0.022)       & (0.021)       & (0.021)       \\
    $\mathbf{X_2}$       & -0.744$^{***}$& -0.570$^{***}$& -0.413$^{***}$& -0.419$^{***}$\\
             & (0.022)       & (0.021)       & (0.020)       & (0.020)       \\
    DID      & 0.148$^{***}$ & 0.099$^{***}$ & 0.023$^{***}$ & 0.020$^{**}$  \\
             & (0.005)       & (0.005)       & (0.007)       & (0.007)       \\
    Constant & 0.071$^{***}$ & -0.471$^{***}$& -0.002         & 0.319$^{***}$ \\
             & (0.002)       & (0.024)       & (0.006)       & (0.045)       \\
    Controls & $\times$      & \checkmark    & $\times$      & \checkmark    \\
    City     & \checkmark    & \checkmark    & \checkmark    & \checkmark    \\
    Year     & $\times$      & $\times$      & \checkmark    & \checkmark    \\
    $R^2$    & 0.500         & 0.593         & 0.666         & 0.673         \\
    Number   & 4426          & 4426          & 4426          & 4426          \\
    \bottomrule
  \end{tabularx}
\end{table}
\newpage
\begin{table}[htbp]
\caption{The Aggregate Moderating Effect from $\mathbf{X_2}$ on Consumption}
  \label{t7}
  \small
  \renewcommand{\arraystretch}{1.2}
  \centering
  \begin{tabularx}{0.9\textwidth}{l *{4}{>{\centering\arraybackslash}X}}
    \toprule
    & (1) & (2) & (3) & (4) \\
    & Consumption & Consumption & Consumption & Consumption \\
    \midrule
    Tech     & 0.672$^{***}$ & 0.596$^{***}$ & 0.530$^{***}$ & 0.526$^{***}$ \\
             & (0.014)       & (0.014)       & (0.014)       & (0.014)       \\
    $\mathbf{X_2}$       & -0.363$^{***}$& -0.295$^{***}$& -0.239$^{***}$& -0.236$^{***}$\\
             & (0.013)       & (0.013)       & (0.013)       & (0.013)       \\
    DID      & 0.065$^{***}$ & 0.055$^{***}$ & 0.034$^{***}$ & 0.031$^{***}$ \\
             & (0.003)       & (0.003)       & (0.004)       & (0.004)       \\
    Constant & 0.040$^{***}$ & -0.133$^{***}$& 0.008          & 0.295$^{***}$ \\
             & (0.001)       & (0.015)       & (0.004)       & (0.029)       \\
    Controls & $\times$      & \checkmark    & $\times$      & \checkmark    \\
    City     & \checkmark    & \checkmark    & \checkmark    & \checkmark    \\
    Year     & $\times$      & $\times$      & \checkmark    & \checkmark    \\
    $R^2$    & 0.618         & 0.652         & 0.683         & 0.694         \\
    Number   & 4426          & 4426          & 4426          & 4426          \\
    \bottomrule
  \end{tabularx}
\end{table}

\subsection{Heterogeneous Effect Analysis}
\textbf{\textit{The Heterogeneity in Technological Progress}}. Regarding the inclusive technological progress $Digital$: In Table 10, we find that the positive impact of this type of technological progress on the social capital stock is marginally increasing. That is, a high level of $Digital$ boosts $Capital$ more strongly than a low level of $Digital$ does. Therefore, we can confirm Hypothesis 1. Furthermore, we find that the positive impact of this type of technological progress on social consumption is also marginally increasing. This indicates that there is still significant potential for $Digital$ to boost $Consumption$ growth. In other words, this type of technological progress has not yet reached a threshold level, continuing to encourage inclusive technological progress will benefit the persistent growth of social consumption.

\begin{table}[htbp]
\caption{The Analysis of Heterogeneity from Technological Progress in Two Types of Digital}
  \label{t10}
  \small
  \renewcommand{\arraystretch}{1.2}
  \centering
  \begin{tabularx}{0.9\textwidth}{l *{4}{>{\centering\arraybackslash}X}}
    \toprule
    & (1) & (2) & (3) & (4) \\
    & $Digital \leqslant 0.28$ & $Digital > 0.28$ & $Digital \leqslant 0.28$ & $Digital > 0.28$ \\
    & Capital & Capital & Consumption & Consumption \\
    \midrule
    Digital  & 0.313$^{**}$  & 0.401$^{***}$ & 0.112$^{*}$   & 0.728$^{***}$ \\
             & (0.104)       & (0.088)       & (0.053)       & (0.078)       \\
    Constant & 0.099         & 0.001         & 0.047         & 0.067         \\
             & (0.065)       & (0.088)       & (0.033)       & (0.078)       \\
    Controls & \checkmark    & \checkmark    & \checkmark    & \checkmark    \\
    City     & \checkmark    & \checkmark    & \checkmark    & \checkmark    \\
    Year     & \checkmark    & \checkmark    & \checkmark    & \checkmark    \\
    $R^2$    & 0.511         & 0.568         & 0.422         & 0.480         \\
    Number   & 2073          & 2353          & 2073          & 2353          \\
    \bottomrule
  \end{tabularx}
\end{table}
\newpage
\begin{table}[htbp]
\caption{The Analysis of Heterogeneity from Technological Progress in Two Types of Tech}
  \label{t12}
  \small
  \renewcommand{\arraystretch}{1.2}
  \centering
  \begin{tabularx}{0.9\textwidth}{l *{4}{>{\centering\arraybackslash}X}}
    \toprule
    & (1) & (2) & (3) & (4) \\
    & $Tech \leqslant 0.005$ & $Tech > 0.005$ & $Tech \leqslant 0.005$ & $Tech > 0.005$ \\
    & Capital & Capital & Consumption & Consumption \\
    \midrule
    Tech     & 4.006$^{***}$ & 0.126$^{***}$ & 2.688$^{***}$ & 0.225$^{***}$ \\
             & (0.392)       & (0.009)       & (0.149)       & (0.006)       \\
    Constant & -0.140$^{***}$& -0.235$^{**}$ & -0.053$^{***}$& 0.426$^{***}$ \\
             & (0.029)       & (0.085)       & (0.011)       & (0.058)       \\
    Controls & \checkmark    & \checkmark    & \checkmark    & \checkmark    \\
    City     & \checkmark    & \checkmark    & \checkmark    & \checkmark    \\
    Year     & \checkmark    & \checkmark    & \checkmark    & \checkmark    \\
    $R^2$    & 0.786         & 0.704         & 0.781         & 0.744         \\
    Number   & 2067          & 2359          & 2067          & 2359          \\
    \bottomrule
  \end{tabularx}
\end{table}

Regarding the specialized technological progress $Tech$: In Table 11, we find that the positive impact of this type of technological progress on the social capital stock exhibits a threshold effect, that is, the enhancement of $Capital$ from high-level $Tech$ is weaker than that from low-level $Tech$. This may indicate that high levels of specialized technological progress are subject to a stronger negative moderating effect from dataization policy than low levels of specialized technological progress. Furthermore, we observe that the positive impact of such technological progress on social consumption follows an inverted U-shaped pattern, which is consistent with the results in Figure 3. Therefore, we can verify Hypothesis 2.

Table 12 and Table 13 present the heterogeneity analysis of the moderating effects for dataization policy in two types of technological progress. 
\begin{table}[htbp]
\caption{The Analysis of Heterogeneity from $\mathbf{X_1}$ in Two Types of Digital}
  \label{t11}
  \small
  \renewcommand{\arraystretch}{1.2}
  \centering
  \begin{tabularx}{0.9\textwidth}{l *{4}{>{\centering\arraybackslash}X}}
    \toprule
    & (1) & (2) & (3) & (4) \\
    & $Digital \leqslant 0.28$ & $Digital > 0.28$ & $Digital \leqslant 0.28$ & $Digital > 0.28$ \\
    & Capital & Capital & Consumption & Consumption \\
    \midrule
    Digital  & 0.138         & 0.192         & 0.047         & 0.082         \\
             & (0.103)       & (0.099)       & (0.053)       & (0.082)       \\
    $\mathbf{X_1}$       & 1.078$^{***}$ & 0.339$^{***}$ & 0.403$^{***}$ & 1.015$^{***}$ \\
             & (0.119)       & (0.072)       & (0.061)       & (0.060)       \\
    DID      & -0.224$^{***}$& -0.149$^{***}$& -0.097$^{***}$& -0.402$^{***}$\\
             & (0.031)       & (0.032)       & (0.016)       & (0.026)       \\
    Constant & 0.070         & 0.067         & 0.041         & 0.259$^{***}$ \\
             & (0.064)       & (0.089)       & (0.033)       & (0.074)       \\
    Controls & \checkmark    & \checkmark    & \checkmark    & \checkmark    \\
    City     & \checkmark    & \checkmark    & \checkmark    & \checkmark    \\
    Year     & \checkmark    & \checkmark    & \checkmark    & \checkmark    \\
    $R^2$    & 0.532         & 0.572         & 0.435         & 0.546         \\
    Number   & 2073          & 2353          & 2073          & 2353          \\
    \bottomrule
  \end{tabularx}
\end{table}
\newpage
\begin{table}[htbp]
\caption{The Analysis of Heterogeneity from $\mathbf{X_2}$ in Two Types of Tech}
  \label{t13}
  \small
  \renewcommand{\arraystretch}{1.2}
  \centering
  \begin{tabularx}{0.9\textwidth}{l *{4}{>{\centering\arraybackslash}X}}
    \toprule
    & (1) & (2) & (3) & (4) \\
    & $Tech \leqslant 0.005$ & $Tech > 0.005$ & $Tech \leqslant 0.005$ & $Tech > 0.005$ \\
    & Capital & Capital & Consumption & Consumption \\
    \midrule
    Tech     & 4.037$^{***}$ & 0.286$^{***}$ & 2.662$^{***}$ & 0.303$^{***}$ \\
             & (0.393)       & (0.028)       & (0.149)       & (0.019)       \\
    $\mathbf{X_2}$       & 1.489         & -0.147$^{***}$& 4.904$^{***}$ & -0.073$^{***}$\\
             & (3.139)       & (0.025)       & (1.186)       & (0.017)       \\
    DID      & 0.008         & -0.017        & -0.007        & 0.006         \\
             & (0.011)       & (0.009)       & (0.004)       & (0.006)       \\
    Constant & -0.135$^{***}$& -0.177$^{*}$  & -0.053$^{***}$& 0.445$^{***}$ \\
             & (0.030)       & (0.084)       & (0.011)       & (0.058)       \\
    Controls & \checkmark    & \checkmark    & \checkmark    & \checkmark    \\
    City     & \checkmark    & \checkmark    & \checkmark    & \checkmark    \\
    Year     & \checkmark    & \checkmark    & \checkmark    & \checkmark    \\
    $R^2$    & 0.787         & 0.711         & 0.786         & 0.746         \\
    Number   & 2067          & 2359          & 2067          & 2359          \\
    \bottomrule
  \end{tabularx}
\end{table}

We find that for inclusive technological progress $Digital$, the moderating effects of dataization policy are consistently positive, indicating that this policy can amplify the effects of such technological progress in promoting the growth of social capital stock and consumption. Therefore, we can conclude that for inclusive technological progress, policy not only increases the proportion of dataization but also promotes the growth of inclusive technological progress, thus amplifying the enhancement of capital stock and consumption. That is: The intensity of the policy’s impact on dataization is weaker than its impact on inclusive technological progress. Consequently, under these conditions, the shock of dataization policy can amplify the positive effects of inclusive technological progress on capital stock and consumption, which is consistent with Hypothesis 3.

However, regarding specialized technological progress $Tech$, we can only conclude that when this type of technological progress is at a high level, the moderating effect of dataization policy on such technological progress is negative, which is consistent with the results in Figure 2 and Figure 3. This also explains why high levels of $Tech$ have a weaker impact on $Capital$ than low levels of $Tech$. This is because high-level specialized technological progress is subject to a significant negative moderating effect from dataization policy, while low-level specialized technological progress is not affected by this effect. Therefore, we can conclude that the policy’s impact on dataization is stronger than its impact on high-level specialized technological progress. Under these conditions, the shock from dataization policy weakens the contribution of high-level specialized technological progress to capital stock and consumption.

\noindent\textbf{\textit{The Heterogeneity in Income and Consumption Structure}}. Now, we analyze the impact of dataization policy on social capital stock ($Capital$) and consumption ($Consumption$) in two types of income and consumption structures according to the Gini coefficient and the Engel coefficient.

Figure 10 illustrates the dynamic effects of the dataization policy shock on social capital stock in two types of income and consumption structures. Figure 11 illustrates the dynamic effects of the dataization policy shock on social consumption in two types of income and consumption structures.
\begin{figure}[htbp]
\centering
\includegraphics[width=15.5cm]{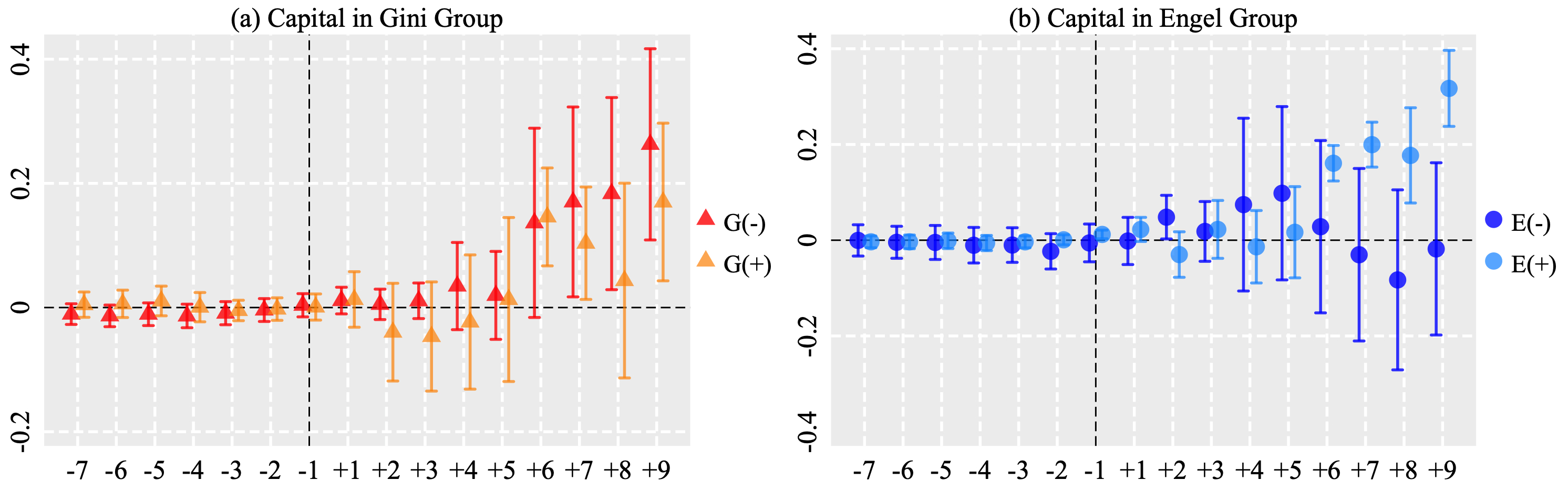}
\caption{\label{fig:A6}The Dynamic Effect of Policy on Capital in Heterogeneous Groups}
\end{figure}
\vspace{-0.4cm}
\begin{figure}[htbp]
\centering
\includegraphics[width=15.5cm]{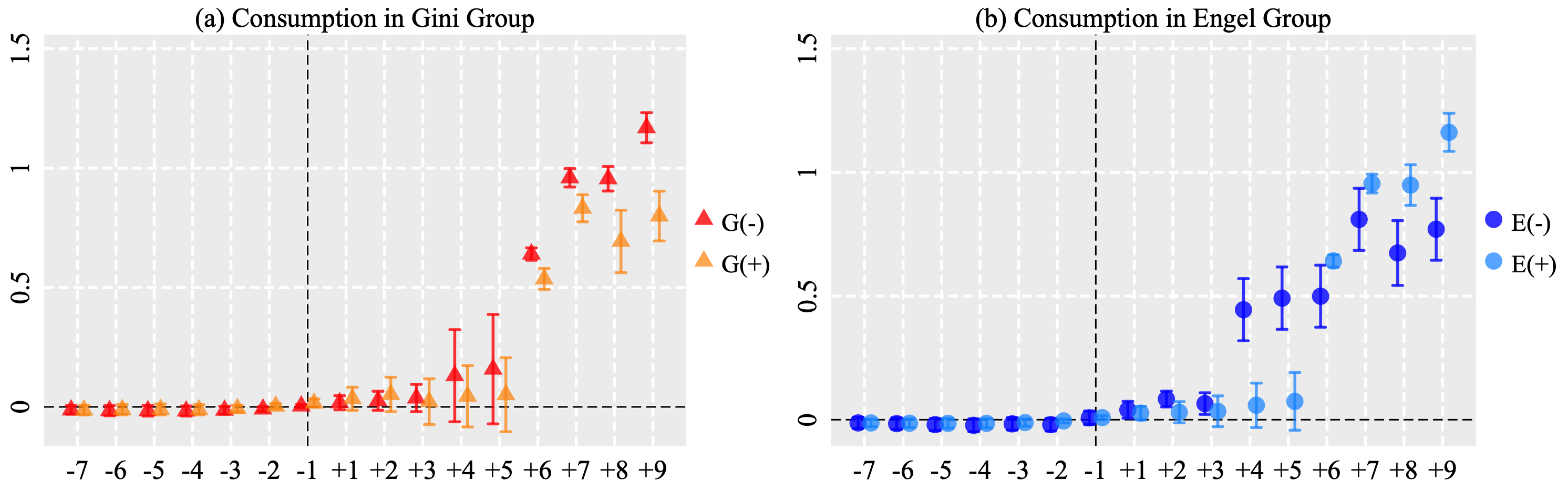}
\caption{\label{fig:A7}The Dynamic Effect of Policy on Consumption in Heterogeneous Groups}
\end{figure}

Based on Figure 10 and Figure 11, we find that in both income groups ($Gini\ Group$), dataization policy delivers significant positive impacts on social capital stock and consumption. However, dataization policy has a stronger effect on boosting social capital stock and consumption in regions with low income inequality ($\mathrm{\textbf{G}}(-)$). Then, in both types of consumption structures ($Engel\ Group$), dataization policy can significantly promote the growth of social capital stock and consumption in regions with high consumption inequality ($\mathrm{\textbf{E}}(+)$), and for social consumption, the positive effects of the policy in regions with high consumption inequality ($\mathrm{\textbf{E}}(+)$) are stronger than those in regions with low consumption inequality ($\mathrm{\textbf{E}}(-)$). However, we can not identify a significant impact of dataization policy on social capital stock in regions with low consumption inequality ($\mathrm{\textbf{E}}(-)$).

Next, we will respectively treat social capital stock ($Capital$) and social consumption ($Consumption$) as control variables in the DID model to construct a stationary general equilibrium, and then we will evaluate the impact of dataization policy shock on technological progress in terms of stationary social capital stock and stationary consumption.

Figure 12 and Figure 13 present the dynamic effects of dataization policy on inclusive technological progress ($Digital$) in stationary general equilibrium. In Figure 12, we find that whether it is in stationary capital stock or stationary consumption, the effect of dataization policy on promoting inclusive technological progress is more significant and stronger in regions with low income inequality ($\mathrm{\textbf{G}}(-)$) than in regions with high income inequality ($\mathrm{\textbf{G}}(+)$). In Figure 13, we find that when it is in stationary capital stock, the effect of dataization policy on promoting inclusive technological progress is stronger in regions with high consumption inequality ($\mathrm{\textbf{E}}(+)$) than in regions with low consumption inequality ($\mathrm{\textbf{E}}(-)$). However, when it is in stationary consumption, the impact of dataization policy is weak. We can only identify a significant positive policy impact in regions with low consumption inequality ($\mathrm{\textbf{E}}(-)$), but the persistence is very fleeting.
\begin{figure}[htbp]
\centering
\includegraphics[width=15.5cm]{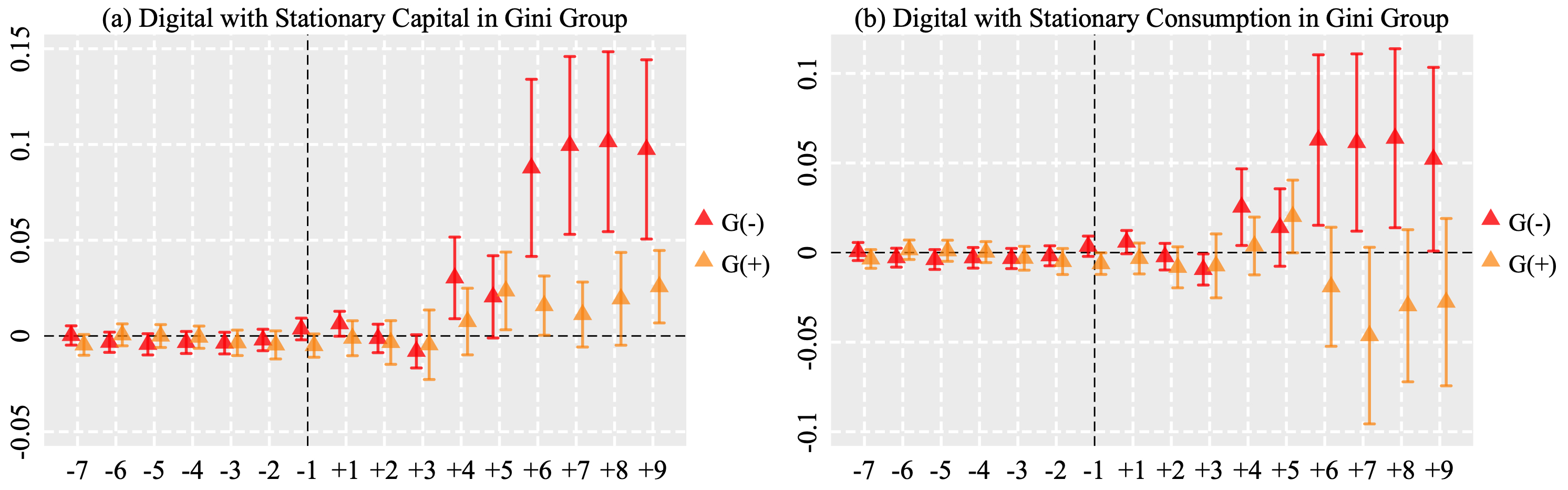}
\caption{\label{fig:A8}The Effect of Policy on Digital with Stationary State in Gini Group}
\end{figure}
\vspace{-0.4cm}
\begin{figure}[htbp]
\centering
\includegraphics[width=15.5cm]{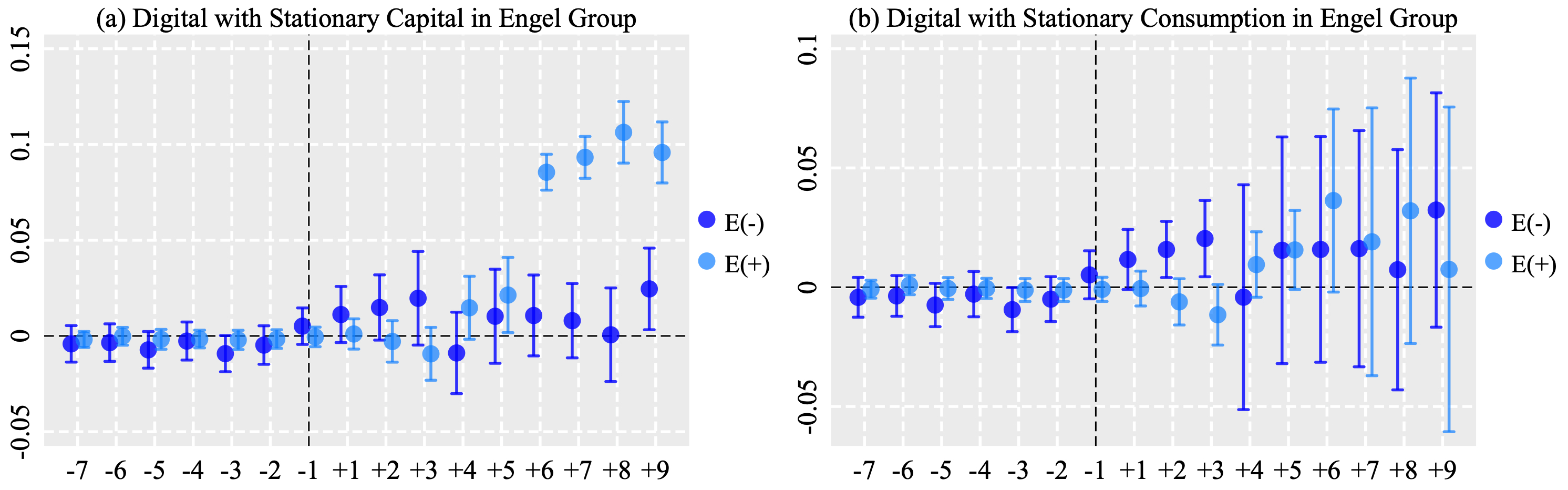}
\caption{\label{fig:A9}The Effect of Policy on Digital with Stationary State in Engel Group}
\end{figure}

Figure 14 and Figure 15 illustrate the dynamic effects of dataization policy shock on specialized technological progress ($Tech$) in stationary general equilibrium. We find that, under the condition of stationary social capital stock, dataization policy can significantly promote the stable growth of specialized technological progress in both types of income structures and consumption structures. Furthermore, compared to the regions with high income inequality ($\mathrm{\textbf{G}}(+)$) and low consumption inequality ($\mathrm{\textbf{E}}(-)$), the effects of dataization policy are stronger in regions with low income inequality ($\mathrm{\textbf{G}}(-)$) and high consumption inequality ($\mathrm{\textbf{E}}(+)$).

Meanwhile, under the condition of stationary social consumption, we find that the contribution of dataization policy to specialized technological progress is limited. In the $Gini\ Group$ shown in Figure 14-b, dataization policy can promote specialized technological progress in regions with low income inequality ($\mathrm{\textbf{G}}(-)$) after a lag period. However, for the regions with high income inequality ($\mathrm{\textbf{G}}(+)$), dataization policy is effective in the early stages of implementation but lose their positive impact on specialized technological progress in the later stages. This is consistent with our Hypothesis 5. In particular, this pattern is more pronounced in the $Engel\ Group$. As shown in Figure 15-b, for both regions with high consumption inequality ($\mathrm{\textbf{E}}(+)$) and low consumption inequality ($\mathrm{\textbf{E}}(-)$), dataization policy is effective in the early stages of implementation but lose their positive impact on specialized technological progress in the later stages.
\begin{figure}[htbp]
\centering
\includegraphics[width=15.5cm]{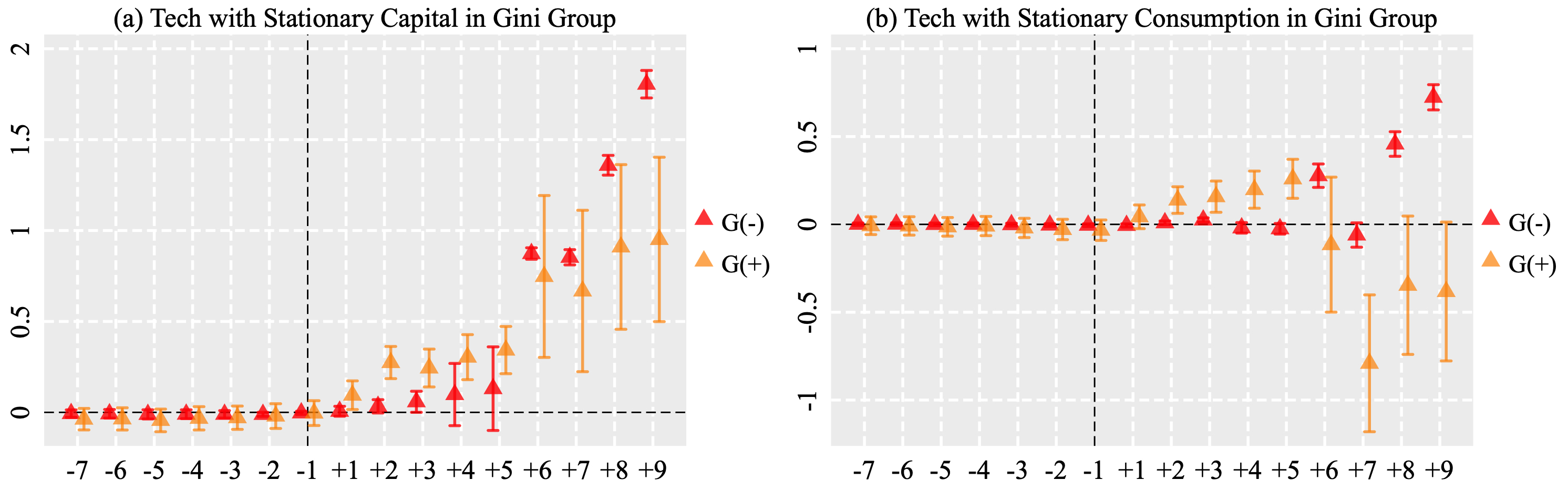}
\caption{\label{fig:A10}The Effect of Policy on Tech with Stationary State in Gini Group}
\end{figure}
\vspace{-0.4cm}
\begin{figure}[htbp]
\centering
\includegraphics[width=15.5cm]{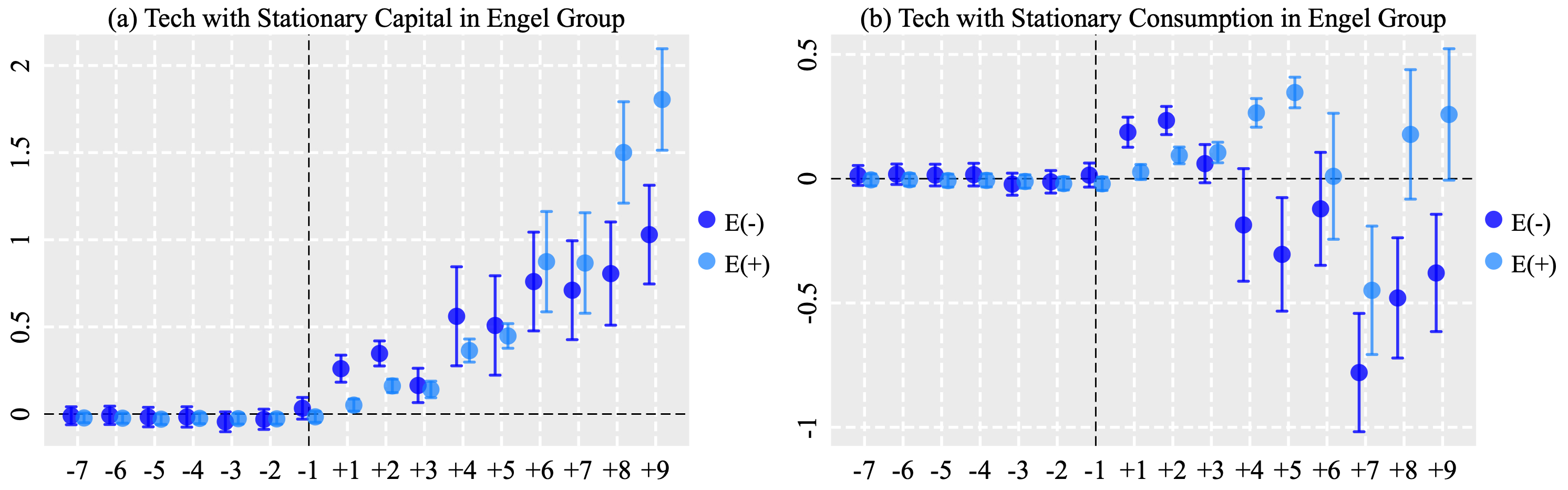}
\caption{\label{fig:A11}The Effect of Policy on Tech with Stationary State in Engel Group}
\end{figure}
\vspace{-0.3cm}

\section{A Partial Equilibrium in Continuous-Time Framework}
\subsection{Analytical Theory}
In this section, based on previous studies, we will try to give an extended discussion to further  analytically explore something about the data distribution of data economy in a continuous-time framework.

We know that firms obtain data from the dataization of aggregate production according to the policy. Here, we assume that firms’ production activities and their responses to the policy are positive, that is, at each stage, firms not only generate positive output but also choose to implement dataization policy to acquire and accumulate data. Therefore, there are two ways for firms to utilize data to promote technological progress: The first comes from historical accumulation, and we define this type of data as data capital $x_t^d$. The second comes from the dataization of current production, which we define as the data stream $y_t^d$. We set that firms will simultaneously choose data capital and data stream as data production factors to improve technology and output, therefore, technological progress in our data economy is the joint distribution of data capital $x_t^d$ and data stream $y_t^d$. Furthermore, firms derive utility from consuming data factors, specifically, firms with standard preferences can derive utility from future data consumption $c_t^d$ with a discount rate $\rho\geqslant 0$:
\[\mathbb{E}_0\int_0^\infty e^{-\rho t}u(c_t^d)dt\]

Here, we set the utility function $u(c_t^d)$ as the CRRA utility function with a relative risk aversion coefficient of $\gamma$.

For the dynamics of data capital $x_t^d$, we define:
\[D(x_t^d,y_t^d)=\dot{x}_t^d=y_t^d+\theta_t^dx_t^d-c_t^d(x_t^d,y_t^d)\]

$D(x_t^d, y_t^d)$ is the policy function. $\theta_t^d>0$ represents the rate of data return on data capital. Specifically, data capital is the capital formed through the accumulation and processing of raw data, which typically manifests as technology or factors of production that facilitate higher levels of output. Therefore, data capital itself can generate data as part of aggregate production. Hence, the data obtained by applying data capital to produce output is denoted as $\theta_t^d x_t^d$. Then, $\theta_t^d$ represents the ability of data capital to generate data in aggregate production, a high value of $\theta_t^d$ indicates that data capital is deeply integrated into production and can generate a large amount of data. In this scenario, the value of data capital is high, and vice versa. We assume that $\theta_t^d=\theta^d$ is a constant under the conditions of equilibrium.

Since the capabilities and the industrial structures of all firms in the economy are heterogeneous, the scale of data capital they can possess, and the value of that data capital will be different among heterogeneous firms. Hence, we assume that data capital is constrained by:
\[x_t^d\geqslant x_l^d,x_l^d\in(-\infty,0)\]

Now, we employ a generalized approach to define the data stream $y_t^d$. We assume that a firm's current production 
$y_t^d$, which could be completely transformed by dataization policy and evolves stochastically over time subject to the constraint $\left[y_l^d, y_h^d\right]$, where $y_h^d>y_l^d\geqslant0$. We then define $y_t^d$ as a stochastic process:
\[dy_t^d=\mu^d(y_t^d)dt+\sigma^d(y_t^d)dW_t^d\]

Where $W_t^d$ is a standard Brownian motion. Function $\mu_y^d=\mu^d(y_t^d)$ is the drift term of this stochastic process, and function $\sigma_y^d=\sigma^d(y_t^d)$ is the diffusion term of this stochastic process.

Therefore, the firms' problem of this section could be written as:
\[\max_{\{c_t^d\}}\mathbb{E}_0\int_0^\infty e^{-\rho t}u(c_t^d)dt\]

s.t.:
\[\begin{gathered}\dot{x}_t^d=y_t^d+\theta_t^dx_t^d-c_t^d(x_t^d,y_t^d)\\x_t^d\geqslant x_l^d,x_l^d\in(-\infty,0)\\dy_t^d=\mu^d(y_t^d)dt+\sigma^d(y_t^d)dW_t^d\end{gathered}\]

Next, we apply the Mean Field Game (MFG) to solve the economic system in partial equilibrium. We define the state space as $\Omega=\left[x_l^d,\infty\right)\times\left[y_l^d,y_h^d\right]$. We denote the stationary value function of our data economy as $v(x^d, y^d)$, and the stationary probability density function of our data distribution as $p(x^d, y^d)$. Hence, the partial equilibrium system to be solved could be written as\footnote{According to MFG, the first equation is the HJB equation, the second is the KFE, the third is the condition for equilibrium probability density, and the fourth is the condition for equilibrium data capital stock.}:
\[\begin{gathered}\rho v(x^d,y^d)=\max_{\{c^d\}}\left[u(c^d)+\frac{\partial v(x^d,y^d)}{\partial x^d}D(x^d,y^d)+\frac{\partial v(x^d,y^d)}{\partial y^d}(\mu_y^d)+\frac{1}{2}\frac{\partial^2v(x^d,y^d)}{\partial(y^d)^2}(\sigma_y^d)^2\right]\\0=-D(x^d,y^d)\frac{\partial p(x^d,y^d)}{\partial x^d}-(\mu_y^d)\frac{\partial p(x^d,y^d)}{\partial y^d}+\frac{1}{2}\left(\sigma_y^d\right)^2\frac{\partial^2p(x^d,y^d)}{\partial(y^d)^2}\\1=\iint_\Omega p(x^d,y^d)d\Omega=\int_{y_l^d}^{y_h^d}\int_{x_l^d}^\infty p(x^d,y^d)dx^ddy^d\\0=\iint_\Omega x^dp(x^d,y^d)d\Omega=\int_{y_l^d}^{y_h^d}\int_{x_l^d}^\infty x^dp(x^d,y^d)dx^ddy^d\end{gathered}\]

It is subject to:
\[\begin{gathered}
\frac{\partial v(x_l^d,y^d)}{\partial x^d}\geqslant\frac{\partial u(y^d+\theta^dx_l^d)}{\partial c^d}\\\frac{\partial v(x^d,y_h^d)}{\partial y^d}=0,\frac{\partial v(x^d,y_l^d)}{\partial y^d}=0
\end{gathered}\]

\subsection{Numerical Simulation}
In this part, we will follow the approach proposed by \cite{achdou2022income} to attempt to find a numerical solution for the partial equilibrium system based on the MFG. We then set specific parameters to compute the probability density function $p(x^d,y^d)$ and perform a numerical simulation to model the distribution of data in our data economy.

Here, we define $y_t^i=\ln y_t^d$:
\[dy_t^i=\kappa_d^i(\tilde{y}-y_t^i)dt+\sigma_d^idW_t^d\]

This is an \textit{Ornstein-Uhlenbeck} (OU) process. For an OU process in a continuous-time framework with time interval $\Delta t$, its autocorrelation coefficient $o_t(\Delta t)$ can be calculated and written as:
\[o_t(\Delta t)=e^{-\kappa_d^i\Delta t}\]

According to the algorithm, we set $\Delta t=1$ and $o_t(\Delta t)=o_d^i\in(0, 1)$, then:
\[\kappa_d^i=-\ln o_d^i\]

Regarding the parameters used in this section: We assume that there are two types of production processes in the data economy that can generate data stream after dataization. The first type is a stable production process $y_t^-$, which returns to the mean $\tilde{y}$ quickly and with low fluctuation. The parameters of this stochastic process are $(c_d^-, \sigma_d^-)=(0.1, 0.1)$. In this scenario, the probability density function of the stationary data distribution is denoted by $\underline{p}(x^-,y^-)$. The second type is the unstable production process $y_t^+$, which returns to the mean $\tilde{y}$ slowly and with high fluctuation. The parameters of this stochastic process are $(c_d^+, \sigma_d^+) = (0.5, 0.5)$. And in this scenario, the probability density function of the stationary data distribution is denoted by $\overline{p}(x^+,y^+)$.

The mean of production is $\tilde{y}=1$. The relative risk aversion coefficient is $\gamma=2$. The rate of data return on data capital is $\theta^d=0.04$. The discount rate is $\rho=0.05$. For details on the methodology and specific algorithms, see \cite{achdou2022income} cited in this section. Numerical simulation results are shown in Figure 16 and Figure 17.
\begin{figure}[htbp]
\centering
\includegraphics[width=15.5cm]{C1.jpg}
\caption{\label{fig:A16}The Probability Density $\underline{p}(x^-,y^-)$ of Stationary Data Distribution}
\end{figure}
\vspace{-0.3cm}
\begin{figure}[htbp]
\centering
\includegraphics[width=15.5cm]{C2.jpg}
\caption{\label{fig:A17}The Probability Density $\overline{p}(x^+,y^+)$ of Stationary Data Distribution}
\end{figure}

Based on these two figures, we find that for stable current production $y_t^-$, its stationary data distribution is a thin-tailed distribution with a high peak. However, for unstable current production $y_t^+$, its stationary data distribution is a thick-tailed distribution with a low peak. Therefore, we can conclude that in the data economy discussed in this paper, data distributions exhibit a peak regardless of whether the production process is stable or unstable. This also implies that neither the data capital derived from established technologies and production factors nor the data streams directly derived from the dataization of total output are linear functions of the data.

In other words: In the data economy under partial equilibrium, there are not always large amounts of high (low) values data capital and high (low) levels data streams. We assume that the economy contains data capital of all possible values $x_t^d\in[x_l^d, x_h^d]$ and data streams of all possible levels $y_t^d\in[y_l^d, y_h^d]$, which we denote as $\mathbb{D}_t=\{x_t^d, y_t^d\}$. When the economy reaches a stationary state, the most common combination with the highest probability density in the economy is $\mathbb{D}_m=\{x_m^d,y_m^d\}$, where $x_l^d<x_m^d<x_h^d$ and $y_l^d<y_m^d<y_h^d$.

To some extent, this provides a deeper explanation of the nexus between dataization and technological progress discussed in our paper. In partial equilibrium, the most prevalent data types in the data economy are not necessarily the most conducive to economic development. Under this circumstance, if we continue to increase the dataization proportion to expand the scale of data in the economy, we may merely change the data at the tails of the data distribution without affecting the most prevalent data types in the data economy. For example, after the dataization of high level production, this portion of data will enter the right tail of the data distribution. However, when we utilize this data again, we can only access the portion with a higher probability density in the data distribution. In this scenario, the effectiveness of dataization policy will experience a period of stagnation: Once the dataization proportion exceeds a certain threshold, further increases in dataization will fail to promote technological progress or other forms of economic development.

\section{Conclusion}
In this paper, we attempt to construct an analytical model to explain the nexus between dataization and technological progress in general equilibrium of macroeconomics with the empirical evidences from Chinese realities.

We first analyze the role of data in aggregate production from the perspective of factor substitution. We find that when production utilizes data elements generated from the dataization of total output, there is a possibility of increasing marginal productivity of capital and labor in aggregate production, respectively. Furthermore, when an incomplete substitution relationship exists between data and traditional production factors, the effect of dataization in promoting technological progress is limited. Specifically, the higher the proportion of data substituting traditional production factors, the more sustainable the effect of dataization in promoting technological progress becomes.

Then, we solve the problems of firms and households based on a production function with data and an endogenous interest rate. We further discuss the impact of dataization and technological progress on general equilibrium. We find that dataization has a negative moderating effect on the impact of technological progress on general equilibrium. Only when policies simultaneously encourage an increase in the dataization proportion and promote technological progress can stable growth in the equilibrium capital stock and equilibrium consumption be achieved. Furthermore, we also find that when equilibrium capital stock is in a stationary state, dataization can promote high levels of technological progress. However, when equilibrium consumption is in a stationary state, the relationship between dataization and technological progress resembles an inverted U-shape. Specifically, increasing the proportion of dataization can boost technological progress at low levels while reducing it at high levels.

Our empirical analysis utilizes macroeconomic data and policy from Chinese cities between 2000 and 2021. We find that the empirical results are largely consistent with the theories proposed in this paper: Chinese dataization policy has achieved both an increase in the dataization proportion and an enhancement of general technological progress, therefore, dataization policy can effectively increase the social capital stock and consumption. Furthermore, under stationary conditions of the social capital stock, dataization policy can sustainably promote technological progress. However, in stationary social consumption, the positive effects of dataization policy on technological progress are limited, in the later stages of policy implementation, the policy loses its positive impact on technological progress. This finding is highly consistent with our theoretical model. We also find that in the current data economy, compared to specialized technological progress, dataization policy not only promotes inclusive technological progress but also exerts a positive moderating effect on the impact of such technological progress on social capital stock and consumption. Therefore, there still exists significant economic potential between dataization policy and inclusive technological progress. We recommend that future policy focus should be shifted toward this type of technological progress, prioritizing the formulation of policies that leverage the advantages of inclusive technological progress to drive economic development.

We also evaluate the effectiveness of the policy based on regional heterogeneity. We find that the impact of the dataization policy on social capital stock and consumption is more robust and stronger in regions with low income inequality and high consumption inequality. Therefore, in the future, we can further refine relevant policies and programs based on the actual conditions of regions with high income inequality and low consumption inequality, thus enhancing the universality of existing dataization policies. At the same time, the results of the heterogeneity tests show that the impact of dataization policy on technological progress is largely consistent with our analysis in the aggregate regression, demonstrating the robustness of our conclusions.

Finally, we study the data distribution of partial equilibrium in a continuous-time framework, providing a deeper explanation of the nexus between dataization and technological progress discussed in our paper. Based on the characteristics of the data distribution, our findings suggest that we should not prioritize obtaining the data from the dataization of firms with very high (low) levels of current production or very high (low) values of data capital, for such data would enter the tails of the data distribution, and the data we can generally obtain and utilize to promote economic development is the portion of the data distribution with a relatively high probability density. That is to say, if a policy focuses solely on dataization at the tails of the data distribution, it may lead to a continuous increase in the proportion of dataization without achieving the expected economic development.

\newpage

\singlespacing
\setlength{\bibsep}{2pt}
\bibliographystyle{plainnat}
\bibliography{main}

\begin{appendices}
\setstretch{1.2}
\section{Econometric Table}\label{Appendix A}

\setcounter{table}{0}
\begin{table}[ht!]
\caption{The Aggregate Regression Effect of Digital on Capital and Consumption}
\label{tab:table1}
\small                             
\renewcommand{\arraystretch}{1} 
\centering
\begin{tabularx}{0.9\textwidth}{l *{4}{>{\centering\arraybackslash}X}}
\toprule
& (1) & (2) & (3) & (4) \\
& Capital & Capital & Consumption & Consumption \\
\midrule
Digital   & $1.251^{***}$ & $0.407^{***}$ & $0.871^{***}$ & $0.632^{***}$ \\
          & (0.037)       & (0.060)       & (0.027)       & (0.046)       \\
GDP       & $0.027^{***}$ & $0.010^{*}$   & $0.015^{***}$ & $0.011^{**}$  \\
          & (0.005)       & (0.005)       & (0.003)       & (0.004)       \\
Company   & 0.012         & $-0.022^{*}$  & $-0.058^{***}$& $-0.066^{***}$\\
          & (0.010)       & (0.011)       & (0.007)       & (0.008)       \\
Population& 0.001         & $-0.077^{***}$& $0.013^{***}$ & -0.010        \\
          & (0.004)       & (0.016)       & (0.003)       & (0.012)       \\
Salary    & $0.011^{***}$ & $0.005^{*}$   & $0.007^{***}$ & $0.004^{*}$   \\
          & (0.002)       & (0.002)       & (0.002)       & (0.002)       \\
Foreign   & $-0.005^{***}$& $-0.011^{***}$& $-0.006^{***}$& $-0.013^{***}$\\
          & (0.001)       & (0.002)       & (0.001)       & (0.002)       \\
Teleservice& $-0.004^{*}$ & -0.003        & $-0.003^{**}$ & $-0.006^{*}$  \\
          & (0.001)       & (0.003)       & (0.001)       & (0.003)       \\
Constant  & $-0.417^{***}$& $0.147^{**}$  & $-0.114^{***}$& 0.062         \\
          & (0.027)       & (0.054)       & (0.020)       & (0.041)       \\
\midrule
City      & \checkmark    & \checkmark    & \checkmark    & \checkmark    \\
Year      & $\times$      & \checkmark    & $\times$      & \checkmark    \\
$R^{2}$   & 0.508         & 0.555         & 0.401         & 0.414         \\
Number    & 4426          & 4426          & 4426          & 4426          \\
\bottomrule
\end{tabularx}
\end{table}

\begin{table}[ht!]
\caption{The Aggregate Regression Effect of Tech on Capital and Consumption}
\label{tab:table2}
\small                           
\renewcommand{\arraystretch}{1}
\centering
\begin{tabularx}{0.9\textwidth}{l *{4}{>{\centering\arraybackslash}X}}
\toprule
& (1) & (2) & (3) & (4) \\
& Capital & Capital & Consumption & Consumption \\
\midrule
Tech      & $0.290^{***}$ & $0.243^{***}$ & $0.312^{***}$ & $0.290^{***}$ \\
          & (0.009)       & (0.008)       & (0.005)       & (0.005)       \\
GDP       & $0.076^{***}$ & 0.007         & $0.043^{***}$ & $0.008^{**}$  \\
          & (0.004)       & (0.004)       & (0.003)       & (0.003)       \\
Company   & $0.125^{***}$ & -0.015        & $0.018^{**}$  & $-0.061^{***}$\\
          & (0.009)       & (0.010)       & (0.005)       & (0.006)       \\
Population& $-0.034^{***}$& $-0.123^{***}$& $-0.014^{***}$& $-0.065^{***}$\\
          & (0.004)       & (0.014)       & (0.003)       & (0.009)       \\
Salary    & $0.023^{***}$ & 0.001         & $0.011^{***}$ & -0.001        \\
          & (0.002)       & (0.002)       & (0.001)       & (0.001)       \\
Foreign   & $-0.003^{*}$  & 0.002         & 0.001         & 0.001         \\
          & (0.001)       & (0.002)       & (0.001)       & (0.001)       \\
Teleservice& $0.006^{***}$& 0.003         & $0.004^{***}$ & 0.001         \\
          & (0.001)       & (0.003)       & (0.001)       & (0.002)       \\
Constant  & $-0.621^{***}$& $0.315^{***}$ & $-0.213^{***}$& $0.298^{***}$ \\
          & (0.025)       & (0.047)       & (0.015)       & (0.030)       \\
\midrule
City      & \checkmark    & \checkmark    & \checkmark    & \checkmark    \\
Year      & $\times$      & \checkmark    & $\times$      & \checkmark    \\
$R^{2}$   & 0.504         & 0.637         & 0.596         & 0.668         \\
Number    & 4426          & 4426          & 4426          & 4426          \\
\bottomrule
\end{tabularx}
\end{table}

\begin{table}[ht!]
\caption{The Aggregate Regression Effect of DID on Capital and Consumption}
\label{tab:table3}
\small                             
\renewcommand{\arraystretch}{1} 
\centering
\begin{tabularx}{0.9\textwidth}{l *{4}{>{\centering\arraybackslash}X}}
\toprule
& (1) & (2) & (3) & (4) \\
& Capital & Capital & Consumption & Consumption \\
\midrule
DID      & $0.096^{***}$ & 0.001         & $0.069^{***}$ & $0.032^{***}$ \\
          & (0.006)       & (0.007)       & (0.005)       & (0.006)       \\
GDP       & $0.074^{***}$ & 0.010         & $0.047^{***}$ & $0.010^{*}$   \\
          & (0.005)       & (0.005)       & (0.004)       & (0.004)       \\
Company   & $0.116^{***}$ & $-0.029^{**}$ & $0.014^{*}$   & $-0.075^{***}$\\
          & (0.010)       & (0.011)       & (0.007)       & (0.008)       \\
Population& $-0.022^{***}$& $-0.076^{***}$& -0.003        & -0.008        \\
          & (0.005)       & (0.016)       & (0.003)       & (0.012)       \\
Salary    & $0.033^{***}$ & $0.006^{*}$   & $0.022^{***}$ & $0.006^{***}$ \\
          & (0.003)       & (0.002)       & (0.002)       & (0.002)       \\
Foreign   & -0.003        & $-0.013^{***}$& $-0.005^{***}$& $-0.016^{***}$\\
          & (0.002)       & (0.002)       & (0.001)       & (0.002)       \\
Teleservice& $0.006^{***}$& -0.003        & $0.003^{**}$  & $-0.006^{*}$  \\
          & (0.002)       & (0.003)       & (0.001)       & (0.003)       \\
Constant  & $-0.667^{***}$& $0.244^{***}$ & $-0.286^{***}$& $0.205^{***}$ \\
          & (0.028)       & (0.052)       & (0.020)       & (0.041)       \\
\midrule
City      & \checkmark    & \checkmark    & \checkmark    & \checkmark    \\
Year      & $\times$      & \checkmark    & $\times$      & \checkmark    \\
$R^{2}$   & 0.406         & 0.550         & 0.292         & 0.393         \\
Number    & 4426          & 4426          & 4426          & 4426          \\
\bottomrule
\end{tabularx}
\end{table}

\begin{table}[ht!]
\caption{The Dataization Shock on Technological Progress with Stationary Capital}
\label{tab:table8}
\small                            
\renewcommand{\arraystretch}{1} 
\centering
\begin{tabularx}{0.9\textwidth}{l *{4}{>{\centering\arraybackslash}X}}
\toprule
& (1) & (2) & (3) & (4) \\
& Digital & Digital & Tech & Tech \\
\midrule
DID      & $0.041^{***}$ & 0.003         & $0.031^{***}$ & $0.085^{***}$ \\
          & (0.002)       & (0.002)       & (0.009)       & (0.012)       \\
Capital   & $0.150^{***}$ & $0.026^{***}$ & $0.703^{***}$ & $0.797^{***}$ \\
          & (0.005)       & (0.004)       & (0.022)       & (0.025)       \\
GDP       & $0.031^{***}$ & -0.001        & -0.012        & -0.001        \\
          & (0.002)       & (0.001)       & (0.007)       & (0.008)       \\
Company   & $0.070^{***}$ & $-0.017^{***}$& $-0.071^{***}$& -0.028        \\
          & (0.003)       & (0.003)       & (0.015)       & (0.018)       \\
Population& $-0.015^{***}$& 0.004         & $0.049^{***}$ & $0.258^{***}$ \\
          & (0.002)       & (0.004)       & (0.007)       & (0.026)       \\
Salary    & $0.013^{***}$ & $0.002^{***}$ & $0.014^{***}$ & $0.017^{***}$ \\
          & (0.001)       & (0.001)       & (0.004)       & (0.004)       \\
Foreign   & $-0.001^{*}$  & $-0.005^{***}$& $-0.031^{***}$& $-0.050^{***}$\\
          & (0.001)       & (0.001)       & (0.003)       & (0.004)       \\
Teleservice& $0.007^{***}$& -0.001        & $-0.006^{**}$ & $-0.020^{***}$\\
          & (0.001)       & (0.001)       & (0.002)       & (0.005)       \\
Constant  & $-0.115^{***}$& $0.233^{***}$ & $0.150^{***}$ & $-0.510^{***}$\\
          & (0.010)       & (0.013)       & (0.043)       & (0.085)       \\
\midrule
City      & \checkmark    & \checkmark    & \checkmark    & \checkmark    \\
Year      & $\times$      & \checkmark    & $\times$      & \checkmark    \\
$R^{2}$   & 0.733         & 0.875         & 0.333         & 0.359         \\
Number    & 4426          & 4426          & 4426          & 4426          \\
\bottomrule
\end{tabularx}
\end{table}

\begin{table}[ht!]
\caption{The Dataization Shock on Technological Progress with Stationary Consumption}
\label{tab:table9}
\small                             
\renewcommand{\arraystretch}{1} 
\centering
\begin{tabularx}{0.9\textwidth}{l *{4}{>{\centering\arraybackslash}X}}
\toprule
& (1) & (2) & (3) & (4) \\
& Digital & Digital & Tech & Tech \\
\midrule
DID      & $0.041^{***}$ & 0.001         & -0.004        & $0.035^{***}$ \\
          & (0.002)       & (0.002)       & (0.008)       & (0.010)       \\
Consumption& $0.196^{***}$& $0.068^{***}$ & $1.471^{***}$ & $1.571^{***}$ \\
          & (0.007)       & (0.005)       & (0.025)       & (0.027)       \\
GDP       & $0.033^{***}$ & -0.002        & $-0.030^{***}$& -0.009        \\
          & (0.002)       & (0.001)       & (0.006)       & (0.007)       \\
Company   & $0.084^{***}$ & $-0.012^{***}$& -0.010        & $0.067^{***}$ \\
          & (0.003)       & (0.003)       & (0.012)       & (0.015)       \\
Population& $-0.018^{***}$& 0.002         & $0.038^{***}$ & $0.210^{***}$ \\
          & (0.002)       & (0.004)       & (0.006)       & (0.021)       \\
Salary    & $0.013^{***}$ & $0.002^{**}$  & 0.005         & $0.012^{***}$ \\
          & (0.001)       & (0.001)       & (0.003)       & (0.003)       \\
Foreign   & -0.001        & $-0.004^{***}$& $-0.027^{***}$& $-0.035^{***}$\\
          & (0.001)       & (0.001)       & (0.002)       & (0.003)       \\
Teleservice& $0.007^{***}$& -0.001        & $-0.007^{***}$& $-0.013^{**}$ \\
          & (0.001)       & (0.001)       & (0.002)       & (0.004)       \\
Constant  & $-0.159^{***}$& $0.225^{***}$ & $0.102^{**}$  & $-0.638^{***}$\\
          & (0.009)       & (0.013)       & (0.034)       & (0.070)       \\
\midrule
City      & \checkmark    & \checkmark    & \checkmark    & \checkmark    \\
Year      & $\times$      & \checkmark    & $\times$      & \checkmark    \\
$R^{2}$   & 0.728         & 0.879         & 0.542         & 0.565         \\
Number    & 4426          & 4426          & 4426          & 4426          \\
\bottomrule
\end{tabularx}
\end{table}

\begin{table}[ht!]
\caption{The Aggregate Moderating Effect from $\mathbf{X_1}$ on Capital}
\label{tab:table4}
\small                            
\renewcommand{\arraystretch}{1}
\centering
\begin{tabularx}{0.9\textwidth}{l *{4}{>{\centering\arraybackslash}X}}
\toprule
& (1) & (2) & (3) & (4) \\
& Capital & Capital & Capital & Capital \\
\midrule
Digital   & $1.382^{***}$ & $1.232^{***}$ & $0.334^{***}$ & $0.303^{***}$ \\
          & (0.028)       & (0.041)       & (0.069)       & (0.069)       \\
$\mathbf{X_1}$        & $-0.162^{***}$& $-0.160^{***}$& $0.170^{***}$ & $0.136^{**}$  \\
          & (0.040)       & (0.042)       & (0.043)       & (0.044)       \\
DID      & $0.095^{***}$ & $0.088^{***}$ & $-0.061^{***}$& $-0.053^{**}$ \\
          & (0.015)       & (0.016)       & (0.018)       & (0.018)       \\
GDP       &               & $0.022^{***}$ &               & $0.010^{*}$   \\
          &               & (0.005)       &               & (0.005)       \\
Company   &               & 0.006         &               & -0.021        \\
          &               & (0.010)       &               & (0.011)       \\
Population&               & 0.002         &               & $-0.078^{***}$\\
          &               & (0.004)       &               & (0.016)       \\
Salary    &               & $0.013^{***}$ &               & 0.004         \\
          &               & (0.002)       &               & (0.002)       \\
Foreign   &               & -0.002        &               & $-0.010^{***}$\\
          &               & (0.002)       &               & (0.002)       \\
Teleservice&              & $-0.004^{*}$  &               & -0.002        \\
          &               & (0.001)       &               & (0.003)       \\
Constant  & $-0.287^{***}$& $-0.405^{***}$& $-0.067^{***}$& $0.174^{**}$  \\
          & (0.008)       & (0.027)       & (0.015)       & (0.055)       \\
\midrule
City      & \checkmark    & \checkmark    & \checkmark    & \checkmark    \\
Year      & $\times$      & $\times$      & \checkmark    & \checkmark    \\
$R^{2}$   & 0.505         & 0.513         & 0.549         & 0.556         \\
Number    & 4426          & 4426          & 4426          & 4426          \\
\bottomrule
\end{tabularx}
\end{table}

\begin{table}[ht!]
\caption{The Aggregate Moderating Effect from $\mathbf{X_1}$ on Consumption}
\label{tab:table5}
\small                             
\renewcommand{\arraystretch}{1} 
\centering
\begin{tabularx}{0.9\textwidth}{l *{4}{>{\centering\arraybackslash}X}}
\toprule
& (1) & (2) & (3) & (4) \\
& Consumption & Consumption & Consumption & Consumption \\
\midrule
Digital   & $0.700^{***}$ & $0.712^{***}$ & $0.317^{***}$ & $0.267^{***}$ \\
          & (0.020)       & (0.030)       & (0.052)       & (0.051)       \\
$\mathbf{X_1}$        & $0.369^{***}$ & $0.336^{***}$ & $0.496^{***}$ & $0.466^{***}$ \\
          & (0.029)       & (0.030)       & (0.032)       & (0.033)       \\
DID      & $-0.105^{***}$& $-0.095^{***}$& $-0.153^{***}$& $-0.147^{***}$\\
          & (0.011)       & (0.012)       & (0.013)       & (0.014)       \\
GDP       &               & $0.018^{***}$ &               & $0.011^{**}$  \\
          &               & (0.003)       &               & (0.004)       \\
Company   &               & $-0.044^{***}$&               & $-0.062^{***}$\\
          &               & (0.007)       &               & (0.008)       \\
Population&               & $0.008^{*}$   &               & -0.014        \\
          &               & (0.003)       &               & (0.012)       \\
Salary    &               & $0.005^{**}$  &               & 0.001         \\
          &               & (0.002)       &               & (0.002)       \\
Foreign   &               & -0.001        &               & $-0.010^{***}$\\
          &               & (0.001)       &               & (0.002)       \\
Teleservice&              & -0.002        &               & -0.004        \\
          &               & (0.001)       &               & (0.002)       \\
Constant  & $-0.138^{***}$& $-0.126^{***}$& $-0.054^{***}$& $0.146^{***}$ \\
          & (0.006)       & (0.019)       & (0.011)       & (0.041)       \\
\midrule
City      & \checkmark    & \checkmark    & \checkmark    & \checkmark    \\
Year      & $\times$      & $\times$      & \checkmark    & \checkmark    \\
$R^{2}$   & 0.413         & 0.423         & 0.430         & 0.445         \\
Number    & 4426          & 4426          & 4426          & 4426          \\
\bottomrule
\end{tabularx}
\end{table}

\begin{table}[ht!]
\caption{The Aggregate Moderating Effect from $\mathbf{X_2}$ on Capital}
\label{tab:table6}
\small                             
\renewcommand{\arraystretch}{0.9}
\centering
\begin{tabularx}{0.9\textwidth}{l *{4}{>{\centering\arraybackslash}X}}
\toprule
& (1) & (2) & (3) & (4) \\
& Capital & Capital & Capital & Capital \\
\midrule
Tech      & $1.039^{***}$ & $0.840^{***}$ & $0.656^{***}$ & $0.665^{***}$ \\
          & (0.023)       & (0.022)       & (0.021)       & (0.021)       \\
$\mathbf{X_2}$        & $-0.744^{***}$& $-0.570^{***}$& $-0.413^{***}$& $-0.419^{***}$\\
          & (0.022)       & (0.021)       & (0.020)       & (0.020)       \\
DID      & $0.148^{***}$ & $0.099^{***}$ & $0.023^{***}$ & $0.020^{**}$  \\
          & (0.005)       & (0.005)       & (0.007)       & (0.007)       \\
GDP       &               & $0.048^{***}$ &               & 0.007         \\
          &               & (0.004)       &               & (0.004)       \\
Company   &               & $0.100^{***}$ &               & -0.004        \\
          &               & (0.008)       &               & (0.009)       \\
Population&               & $-0.026^{***}$&               & $-0.130^{***}$\\
          &               & (0.004)       &               & (0.014)       \\
Salary    &               & $0.017^{***}$ &               & 0.001         \\
          &               & (0.002)       &               & (0.002)       \\
Foreign   &               & $0.004^{**}$  &               & -0.001        \\
          &               & (0.002)       &               & (0.002)       \\
Teleservice&              & $0.005^{***}$ &               & 0.001         \\
          &               & (0.001)       &               & (0.003)       \\
Constant  & $0.071^{***}$ & $-0.471^{***}$& -0.002        & $0.319^{***}$ \\
          & (0.002)       & (0.024)       & (0.006)       & (0.045)       \\
\midrule
City      & \checkmark    & \checkmark    & \checkmark    & \checkmark    \\
Year      & $\times$      & $\times$      & \checkmark    & \checkmark    \\
$R^{2}$   & 0.500         & 0.593         & 0.666         & 0.673         \\
Number    & 4426          & 4426          & 4426          & 4426          \\
\bottomrule
\end{tabularx}
\end{table}

\begin{table}[ht!]
\caption{The Aggregate Moderating Effect from $\mathbf{X_2}$ on Consumption}
\label{tab:table7}
\small                             
\renewcommand{\arraystretch}{1} 
\centering
\begin{tabularx}{0.9\textwidth}{l *{4}{>{\centering\arraybackslash}X}}
\toprule
& (1) & (2) & (3) & (4) \\
& Consumption & Consumption & Consumption & Consumption \\
\midrule
Tech      & $0.672^{***}$ & $0.596^{***}$ & $0.530^{***}$ & $0.526^{***}$ \\
          & (0.014)       & (0.014)       & (0.014)       & (0.014)       \\
$\mathbf{X_2}$        & $-0.363^{***}$& $-0.295^{***}$& $-0.239^{***}$& $-0.236^{***}$\\
          & (0.013)       & (0.013)       & (0.013)       & (0.013)       \\
DID      & $0.065^{***}$ & $0.055^{***}$ & $0.034^{***}$ & $0.031^{***}$ \\
          & (0.003)       & (0.003)       & (0.004)       & (0.004)       \\
GDP       &               & $0.027^{***}$ &               & $0.007^{**}$  \\
          &               & (0.003)       &               & (0.003)       \\
Company   &               & 0.004         &               & $-0.054^{***}$\\
          &               & (0.005)       &               & (0.006)       \\
Population&               & $-0.010^{***}$&               & $-0.068^{***}$\\
          &               & (0.002)       &               & (0.009)       \\
Salary    &               & $0.008^{***}$ &               & -0.001        \\
          &               & (0.001)       &               & (0.001)       \\
Foreign   &               & $0.005^{***}$ &               & 0.001         \\
          &               & (0.001)       &               & (0.001)       \\
Teleservice&              & $0.003^{***}$ &               & -0.001        \\
          &               & (0.001)       &               & (0.002)       \\
Constant  & $0.040^{***}$ & $-0.133^{***}$& 0.008         & $0.295^{***}$ \\
          & (0.001)       & (0.015)       & (0.004)       & (0.029)       \\
\midrule
City      & \checkmark    & \checkmark    & \checkmark    & \checkmark    \\
Year      & $\times$      & $\times$      & \checkmark    & \checkmark    \\
$R^{2}$   & 0.618         & 0.652         & 0.683         & 0.694         \\
Number    & 4426          & 4426          & 4426          & 4426          \\
\bottomrule
\end{tabularx}
\end{table}

\begin{table}[ht!]
\caption{The Analysis of Heterogeneity from Technological Progress in Two Types of Digital}
\label{tab:table10}
\small                           
\renewcommand{\arraystretch}{1} 
\centering
\begin{tabularx}{0.9\textwidth}{l *{4}{>{\centering\arraybackslash}X}}
\toprule
& (1) & (2) & (3) & (4) \\
& \shortstack{Digital $\leqslant$ 0.28 \\ Capital} & \shortstack{Digital $>$ 0.28 \\ Capital} & \shortstack{Digital $\leqslant$ 0.28 \\ Consumption} & \shortstack{Digital $>$ 0.28 \\ Consumption} \\
\midrule
Digital   & $0.313^{**}$  & $0.401^{***}$ & $0.112^{*}$   & $0.728^{***}$ \\
          & (0.104)       & (0.088)       & (0.053)       & (0.078)       \\
GDP       & $0.014^{*}$   & 0.002         & $0.009^{**}$  & $0.011^{*}$   \\
          & (0.006)       & (0.006)       & (0.003)       & (0.005)       \\
Company   & $-0.036^{**}$ & -0.016        & $-0.021^{***}$& $-0.184^{***}$\\
          & (0.012)       & (0.021)       & (0.006)       & (0.019)       \\
Population& -0.030        & $-0.083^{***}$& -0.001        & -0.006        \\
          & (0.021)       & (0.019)       & (0.010)       & (0.017)       \\
Salary    & 0.002         & $0.009^{**}$  & -0.003        & $0.008^{**}$  \\
          & (0.004)       & (0.003)       & (0.002)       & (0.003)       \\
Foreign   & $-0.012^{***}$& $-0.014^{***}$& $-0.005^{**}$ & $-0.017^{***}$\\
          & (0.003)       & (0.003)       & (0.002)       & (0.003)       \\
Teleservice& -0.001       & -0.002        & 0.001         & $-0.008^{*}$  \\
          & (0.005)       & (0.004)       & (0.003)       & (0.003)       \\
Constant  & 0.099         & 0.001         & 0.047         & 0.067         \\
          & (0.065)       & (0.088)       & (0.033)       & (0.078)       \\
\midrule
City      & \checkmark    & \checkmark    & \checkmark    & \checkmark    \\
Year      & \checkmark    & \checkmark    & \checkmark    & \checkmark    \\
$R^{2}$   & 0.511         & 0.568         & 0.422         & 0.480         \\
Number    & 2073          & 2353          & 2073          & 2353          \\
\bottomrule
\end{tabularx}
\end{table}

\begin{table}[ht!]
\caption{The Analysis of Heterogeneity from Technological Progress in Two Types of Tech}
\label{tab:table12}
\small                             
\renewcommand{\arraystretch}{1} 
\centering
\begin{tabularx}{0.9\textwidth}{l *{4}{>{\centering\arraybackslash}X}}
\toprule
& (1) & (2) & (3) & (4) \\
& \shortstack{Tech $\leqslant$ 0.005 \\ Capital} & \shortstack{Tech $>$ 0.005 \\ Capital} & \shortstack{Tech $\leqslant$ 0.005 \\ Consumption} & \shortstack{Tech $>$ 0.005 \\ Consumption} \\
\midrule
Tech      & $4.006^{***}$ & $0.126^{***}$ & $2.688^{***}$ & $0.225^{***}$ \\
          & (0.392)       & (0.009)       & (0.149)       & (0.006)       \\
GDP       & $0.027^{***}$ & 0.007         & -0.001        & $0.010^{**}$  \\
          & (0.003)       & (0.005)       & (0.001)       & (0.004)       \\
Company   & $0.013^{***}$ & $0.103^{***}$ & $0.002^{**}$  & $-0.159^{***}$\\
          & (0.002)       & (0.025)       & (0.001)       & (0.017)       \\
Population& 0.007         & $-0.122^{***}$& $0.021^{***}$ & $-0.065^{***}$\\
          & (0.011)       & (0.018)       & (0.004)       & (0.012)       \\
Salary    & 0.001         & -0.002        & -0.001        & -0.001        \\
          & (0.002)       & (0.003)       & (0.001)       & (0.002)       \\
Foreign   & $0.002^{**}$  & $0.011^{**}$  & 0.001         & $0.009^{**}$  \\
          & (0.001)       & (0.004)       & (0.001)       & (0.003)       \\
Teleservice& 0.001        & $0.009^{*}$   & 0.001         & 0.003         \\
          & (0.001)       & (0.004)       & (0.001)       & (0.003)       \\
Constant  & $-0.140^{***}$& $-0.235^{**}$ & $-0.053^{***}$& $0.426^{***}$ \\
          & (0.029)       & (0.085)       & (0.011)       & (0.058)       \\
\midrule
City      & \checkmark    & \checkmark    & \checkmark    & \checkmark    \\
Year      & \checkmark    & \checkmark    & \checkmark    & \checkmark    \\
$R^{2}$   & 0.786         & 0.704         & 0.781         & 0.744         \\
Number    & 2067          & 2359          & 2067          & 2359          \\
\bottomrule
\end{tabularx}
\end{table}

\begin{table}[ht!]
\caption{The Analysis of Heterogeneity from $\mathbf{X_1}$ in Two Types of Digital}
\label{tab:table11}
\small                            
\renewcommand{\arraystretch}{1} 
\centering
\begin{tabularx}{0.9\textwidth}{l *{4}{>{\centering\arraybackslash}X}}
\toprule
& (1) & (2) & (3) & (4) \\
& \shortstack{Digital $\leqslant$ 0.28 \\ Capital} & \shortstack{Digital $>$ 0.28 \\ Capital} & \shortstack{Digital $\leqslant$ 0.28 \\ Consumption} & \shortstack{Digital $>$ 0.28 \\ Consumption} \\
\midrule
Digital   & 0.138         & 0.192         & 0.047         & 0.082         \\
          & (0.103)       & (0.099)       & (0.053)       & (0.082)       \\
$\mathbf{X_1}$        & $1.078^{***}$ & $0.339^{***}$ & $0.403^{***}$ & $1.015^{***}$ \\
          & (0.119)       & (0.072)       & (0.061)       & (0.060)       \\
DID      & $-0.224^{***}$& $-0.149^{***}$& $-0.097^{***}$& $-0.402^{***}$\\
          & (0.031)       & (0.032)       & (0.016)       & (0.026)       \\
GDP       & $0.013^{*}$   & 0.003         & $0.009^{**}$  & $0.013^{**}$  \\
          & (0.006)       & (0.006)       & (0.003)       & (0.005)       \\
Company   & $-0.038^{***}$& -0.011        & $-0.023^{***}$& $-0.166^{***}$\\
          & (0.011)       & (0.021)       & (0.006)       & (0.018)       \\
Population& -0.028        & $-0.090^{***}$& -0.001        & -0.023        \\
          & (0.020)       & (0.019)       & (0.010)       & (0.016)       \\
Salary    & $0.009^{*}$   & 0.006         & -0.001        & 0.001         \\
          & (0.004)       & (0.003)       & (0.002)       & (0.003)       \\
Foreign   & -0.005        & $-0.013^{***}$& -0.002        & $-0.014^{***}$\\
          & (0.003)       & (0.003)       & (0.002)       & (0.003)       \\
Teleservice& 0.002        & -0.001        & 0.002         & -0.004        \\
          & (0.005)       & (0.004)       & (0.003)       & (0.003)       \\
Constant  & 0.070         & 0.067         & 0.041         & $0.259^{***}$ \\
          & (0.064)       & (0.089)       & (0.033)       & (0.074)       \\
\midrule
City      & \checkmark    & \checkmark    & \checkmark    & \checkmark    \\
Year      & \checkmark    & \checkmark    & \checkmark    & \checkmark    \\
$R^{2}$   & 0.532         & 0.572         & 0.435         & 0.546         \\
Number    & 2073          & 2353          & 2073          & 2353          \\
\bottomrule
\end{tabularx}
\end{table}

\begin{table}[ht!]
\caption{The Analysis of Heterogeneity from $\mathbf{X_2}$ in Two Types of Tech}
\label{tab:table13}
\small                             
\renewcommand{\arraystretch}{1}
\centering
\begin{tabularx}{0.9\textwidth}{l *{4}{>{\centering\arraybackslash}X}}
\toprule
& (1) & (2) & (3) & (4) \\
& \shortstack{Tech $\leqslant$ 0.005 \\ Capital} & \shortstack{Tech $>$ 0.005 \\ Capital} & \shortstack{Tech $\leqslant$ 0.005 \\ Consumption} & \shortstack{Tech $>$ 0.005 \\ Consumption} \\
\midrule
Tech      & $4.037^{***}$ & $0.286^{***}$ & $2.662^{***}$ & $0.303^{***}$ \\
          & (0.393)       & (0.028)       & (0.149)       & (0.019)       \\
$\mathbf{X_2}$        & 1.489         & $-0.147^{***}$& $4.904^{***}$ & $-0.073^{***}$\\
          & (3.139)       & (0.025)       & (1.186)       & (0.017)       \\
DID      & 0.008         & -0.017        & -0.007        & 0.006         \\
          & (0.011)       & (0.009)       & (0.004)       & (0.006)       \\
GDP       & $0.027^{***}$ & 0.008         & -0.000        & $0.010^{**}$  \\
          & (0.003)       & (0.005)       & (0.001)       & (0.004)       \\
Company   & $0.013^{***}$ & $0.094^{***}$ & $0.002^{**}$  & $-0.161^{***}$\\
          & (0.002)       & (0.025)       & (0.001)       & (0.017)       \\
Population& 0.005         & $-0.123^{***}$& $0.022^{***}$ & $-0.065^{***}$\\
          & (0.011)       & (0.017)       & (0.004)       & (0.012)       \\
Salary    & -0.001        & -0.003        & -0.001        & -0.001        \\
          & (0.002)       & (0.003)       & (0.001)       & (0.002)       \\
Foreign   & $0.001^{**}$  & $0.009^{*}$   & 0.001         & $0.008^{**}$  \\
          & (0.001)       & (0.004)       & (0.001)       & (0.003)       \\
Teleservice& 0.001        & $0.008^{*}$   & 0.001         & 0.003         \\
          & (0.001)       & (0.004)       & (0.001)       & (0.003)       \\
Constant  & $-0.135^{***}$& $-0.177^{*}$  & $-0.053^{***}$& $0.445^{***}$ \\
          & (0.030)       & (0.084)       & (0.011)       & (0.058)       \\
\midrule
City      & \checkmark    & \checkmark    & \checkmark    & \checkmark    \\
Year      & \checkmark    & \checkmark    & \checkmark    & \checkmark    \\
$R^{2}$   & 0.787         & 0.711         & 0.786         & 0.746         \\
Number    & 2067          & 2359          & 2067          & 2359          \\
\bottomrule
\end{tabularx}
\end{table}

\end{appendices}

\end{document}